\documentclass[english,notitlepage,nofootinbib]{revtex4-1}
\pdfoutput=1
\usepackage[T1]{fontenc}
\usepackage[latin9]{inputenc}
\usepackage{color}
\usepackage{comment}
\usepackage{float}
\usepackage{graphicx}
\usepackage{wasysym}
\usepackage{esint}
\usepackage{hyperref}
\usepackage[normalem]{ulem}%to use strike out

\newif\ifShowModified

\ifShowModified
\newcommand{\del}[1]{{\color{red} #1}}
\else
\newcommand{\del}[1]{}
\fi

\makeatletter

\providecommand{\tabularnewline}{\\}

\usepackage[percent]{overpic}
\usepackage{afterpage}

\makeatother

\usepackage{babel}
\begin{document}

\title{Neutrino Parameters from Reactor and Accelerator Neutrino
Experiments  %Effective $\theta_13$ and its application in searching CP violation in neutrino mixing
%\\
%or \\
%Effective $\theta_13$ and its application in combined analyses of reactor and accelerator neutrino experimentsor ...\\
}

\author{Manfred Lindner, Werner Rodejohann and Xun-Jie Xu}

\affiliation{\textcolor{black}{Max-Planck-Institut f\"ur Kernphysik, Postfach
103980, D-69029 Heidelberg, Germany}}

\date{\today}
\begin{abstract}\noindent
We revisit correlations of neutrino oscillation parameters in
reactor and long-baseline neutrino oscillation experiments.
%, including both analytical and numerical studies.
A framework based on an  effective value of $\theta_{13}$ is presented, which can be used to analytically study the correlations and explain some questions including why and when $\delta_{CP}$ has the best fit value of $-\pi/2$, why current and future long-baseline experiments will have less precision of $\delta_{CP}$ around $\pm \pi/2$ than that around zero, etc.
%which can easily reproduce experimental results. It also
%clarifies why current and future long-baseline experiments will have less
%precision on values around $\delta_{CP} = \pm \pi/2$ than on values around $\delta_{CP} = 0$.
%Optimization potential for the determination of the theoretically
%very interesting values around $\delta_{CP} = -\pi/2$ is also pointed out, which would require that
%future runs of accelerator experiments
%are not equally shared $1:1$ in neutrino and antineutrino modes, but rather $2:1$.
Recent hints on the CP phase
%and the mass ordering
are then considered
from the point of view that different reactor and long-baseline neutrino
experiments provide currently different best-fit values of
$\theta_{13}$ and $\theta_{23}$. We point out that the significance of the hints changes for the
different available best-fit values.

\end{abstract}
\maketitle

\section{Introduction}

Short-baseline reactor and long-baseline accelerator neutrino experiments are of
huge importance in the era of neutrino oscillation precision measurements. The
precision determination of $\theta_{23}$, and the measurements of the CP phase $\delta_{CP}$ and the
neutrino mass ordering would help further understanding the leptonic flavor sector and could bring various
insights in models behind neutrino mass and lepton mixing.
%If these measurements would succeed, they would
%continue the success story that started with the determination of nonzero $\theta_{13}$
%by T2K \cite{Abe:2011sj}, Double Chooz \cite{Abe:2011fz}, Daya Bay \cite{An:2012eh}
%and RENO \cite{Ahn:2012nd}.
%For global fits of all data, including also the more complicated addition of atmospheric data, see
%\cite{Capozzi:2016rtj,Esteban:2016qun,deSalas:2017kay}.

The interplay of long- and short-baseline experiments is exemplified by the
dependence of the electron antineutrino survival probability in reactor experiments, which depends on
$\theta_{13}$, and the $\nu_{\mu}\rightarrow\nu_{e}$ (or $\bar\nu_{\mu}\rightarrow\bar\nu_{e}$) transition
probability in accelerator experiments, which depends on $\theta_{13}$, $\theta_{23}$, $\delta_{CP}$
and the mass ordering. Inserting the reactor determination of $\theta_{13}$, as well as
$\theta_{23}$ values from atmospheric data or long-baseline muon neutrino survival probabilities, into
$\nu_{\mu}\rightarrow\nu_{e}$ (or $\bar\nu_{\mu}\rightarrow\bar\nu_{e}$) measurements can give hints
on the CP phase \cite{T2K2014PRL,Bian:2015opa}.
Combining $\nu_{\mu}\rightarrow\nu_{e}$ plus $\bar\nu_{\mu}\rightarrow\bar\nu_{e}$ data
with additional input on $\theta_{23}$ is enough to get sensitivity on $\delta_{CP}$ \cite{Abe:2017uxa,Abe:2017vif}.
As matter effects start to play a non-negligible role also some sensitivity on the mass ordering
can occur by combining different data sets.
Indeed, recently hints towards non-trivial values of the CP phase $\delta_{CP}$ and
some preference of the normal ordering over the inverted one were found in combining
short-baseline reactor and long-baseline accelerator
neutrino experiments, see Refs.\
\cite{Capozzi:2016rtj,Esteban:2016qun,deSalas:2017kay} for recent global fits.

In this respect, one should note here the different values of $\theta_{13}$ that have been
determined by the three reactor neutrino experiments:
$\theta_{13}={8.43^{\circ}}_{-0.17}^{+0.17}$ (Daya Bay \cite{An:2016ses}),
${8.62}^{\circ}{}_{-0.57}^{+0.54}$ (RENO \cite{Seo:2016dbz}) and
${{9.73}^{\circ}}_{-0.85}^{+0.79}$ (Double Chooz \cite{Matsubara:2016prb}).
It is possible that the central value of
$\theta_{13}$ is shifted (e.g.\ from a joint analysis of the collaborations), which will
then have consequences on the determination of the other neutrino parameters. Moreover,
the best-fit values (normal ordering for definiteness)
of $\theta_{23}$ are $46.8^{\circ}$ from T2K \cite{Abe:2017bay}
and $39.5^{\circ}$ $(52.2^{\circ})$ from NO$\nu$A \cite{Adamson:2017qqn},
where the latter has two almost equally good best-fit
points\footnote{Most recent data from NO$\nu$A \cite{NOVAlink} does not seem to confirm this feature any longer. Changes of
  such best-fit points are typical for experiments with comparably low
event numbers. For the sake of illustration of the general situation
in the field, we will work with the two best-fit points from \cite{Adamson:2017qqn}.}.
In this paper we try to estimate the possible impact of the different central values of
$\theta_{13}$ and $\theta_{23}$ on the hints for CP violation and the normal mass ordering.
Towards this end, we re-analyze the correlation of long-baseline
$\nu_\mu \to \nu_e$
$(\bar\nu_\mu \to \bar \nu_e)$ transition
and reactor $\bar \nu_e \to \bar \nu_e$ survival oscillations.

We study the impact of combining accelerator neutrino data with antineutrino data,
and with reactor data. We investigate the impact of the different
$\theta_{13}$ and $\theta_{23}$ best-fit values on the
determination of $\delta_{CP}$, as well as on the preference for the normal mass ordering.
We demonstrate that the significance can shift, and should thus be taken with care.
%For instance, the significance of CP violation could be enhanced or reduced
%the preference for the normal mass ordering from appearance data
%may be reduced by more than one standard deviation.
We also remark that the theoretically
very interesting\footnote{See Refs.\
\cite{Harrison:2002et,Ma:2002ce,Babu:2002dz,Ma:2002ge,Grimus:2003yn,Nishi:2013jqa,
  Ma:2013mga,
  Fraser:2014yha,He:2015gba,Ma:2015gka,DiIura:2015kfa,Mohapatra:2015gwa,
  Zhou:2014sya,Joshipura:2015dsa,He:2015xha,He:2012yt,Rodejohann:2017lre,Rodejohann:2015hka,
  Zhao:2017yvw,Nishi:2016wki,Chen:2015siy,
  Fukuyama:2017qxb,Li:2015jxa}
for an incomplete list.} value of $\delta_{CP} = -\pi/2$ can be measured with currently running
accelerator experiments better if future runs
are not equally shared $1:1$ in neutrino and antineutrino modes, but rather $2:1$.  \\

The remainder of this paper is organized as follows.
In Sec.~\ref{sec:basic}, we perform semi-analytical analyses on the interplay of the neutrino parameters, including the impact of
the different available central values of $\theta_{13}$ and $\theta_{23}$ on the CP phase
$\delta_{CP}$ and the mass ordering.
%such as $\theta_{13}$, $\theta_{23}$, $\delta_{CP}$, etc.
In Sec.~\ref{sec:applications}, we further adopt numerical calculations  to quantitatively study the  phenomenology,
which not only verifies our analyses in Sec.~\ref{sec:basic} but also
gives a more precise description.
Finally, we summarize in Sec.~\ref{sec:Conclusion} and leave some detailed issues concerning the semi-analytical analyses in the appendix.

\section{Analytical approach: effective $\theta_{13}$\label{sec:basic}}

We start from the approximate neutrino oscillation formula \cite{Freund:2001pn,Cervera:2000kp,PDG2014}
that can accurately describe $\nu_{\mu}\rightarrow\nu_{e}$ transitions in current
and upcoming  accelerator neutrino experiments such as T2K \cite{T2K2013PRL,T2K2014PRL},
MINOS \cite{MINOS}, NO$\nu$A \cite{NOVA,NOVA2}, DUNE \cite{Acciarri:2015uup} or T2HK \cite{Abe:2014oxa}:
\begin{eqnarray}
P(\nu_{\mu}\rightarrow\nu_{e}) & \approx & 4s_{13}^{2}c_{13}^{2}s_{23}^{2}\frac{\sin^{2}(1-A)\Delta}{(1-A)^{2}}\nonumber \\
 &  & -8\alpha J_{CP}\sin\Delta\frac{\sin A\Delta}{A}\frac{\sin(1-A)\Delta}{1-A}\nonumber \\
 &  & +8\alpha(J_{CP}\cot\delta_{CP})\cos\Delta\frac{\sin A\Delta}{A}\frac{\sin(1-A)\Delta}{1-A}\nonumber \\
 &  & +4\alpha^{2}s_{12}^{2}c_{12}^{2}c_{23}^{2}\frac{\sin^{2}A\Delta}{A^{2}}.\label{eq:eff}
\end{eqnarray}
Here $(s_{ij},\thinspace c_{ij})\equiv(\sin\theta_{ij},\thinspace\cos\theta_{ij})$,
$\alpha\equiv\Delta m_{21}^{2}/\Delta m_{31}^{2}$, $A\equiv2\sqrt{2}G_{F}N_{e}E/\Delta m_{32}^{2}$
($N_{e}$ is the electron number density in matter), $\Delta\equiv\Delta m_{32}^{2}L/(4E)$
and
\begin{equation}
J_{CP}=\frac{1}{8}\sin\delta_{CP}\sin 2\theta_{12}\sin
2\theta_{13}\sin 2\theta_{23} c_{13}.\label{eq:eff-1}
\end{equation}
The matter effect \cite{Wolfenstein:1977ue,Mikheev:1986gs,Mikheev:1986wj}
is included by the parameter $A$. For $\overline{\nu}_{\mu}\rightarrow\overline{\nu}_{e}$
transitions, Eq.~(\ref{eq:eff}) can be used by replacing $\delta_{CP}\rightarrow-\delta_{CP}$
(implying $J_{CP}\rightarrow-J_{CP}$)
and $A\rightarrow-A$. For the two possibilities of the mass ordering,
namely the normal/inverted ordering (NO/IO), the formula allows negative
$\Delta m_{32}^{2}$, i.e.\ in the IO one takes negative values
of $A$, $\alpha$ and $\Delta$.
Eq.~(\ref{eq:eff}) is derived from the series expansion in $\alpha$
where the leading order (LO), next-to-leading order (NLO), and next-to-next-to
leading order (NNLO) terms correspond to the first, second plus third, and
last rows of Eq.~(\ref{eq:eff}),
respectively\footnote{More exactly, the accuracy requires not only small $\alpha$, but
also small $\alpha\Delta$, $s_{13}^{2}$, etc. See Ref.\ \cite{Xu:2015kma}
for a detailed discussion on the validity of Eq.\ (\ref{eq:eff}).}. At  LO, the oscillation probability of $\nu_{\mu}\rightarrow\nu_{e}$
depends on $\theta_{13}$ and $\theta_{23}$, but is independent of $\delta_{CP}$, which appears
at NLO.

Let us first focus on the LO and neglect the higher order terms so
Eq.~(\ref{eq:eff}) can be approximately written as
\begin{equation}
P(\nu_{\mu}\rightarrow\nu_{e})\approx\frac{1}{2}\sin^{2}2\theta_{13}^{{\rm eff}}\frac{\sin^{2}(1-A)\Delta}{(1-A)^{2}},\label{eq:eff-2}
\end{equation}
where
\begin{equation}
\sin^{2}2\theta_{13}^{{\rm eff}}=2s_{23}^{2}\sin^{2}2\theta_{13}+{\cal O}(\alpha).\label{eq:eff-3}
\end{equation}
Here we have introduced an angle $\theta_{13}^{{\rm eff}}$ which we will
refer to as \emph{effective $\theta_{13}$} in this paper. At this
stage it is only the coefficient $4s_{13}^{2}c_{13}^{2}s_{23}^{2}$
of the LO expression, but below we will further generalize it to higher orders and
provide a more general definition. Note that in the limit $\alpha\rightarrow0$
and $\theta_{23}\rightarrow45^{\circ}$, $\theta_{13}^{{\rm eff}}$
is equal to $\theta_{13}$. At this level a correlation exists only
with $\theta_{23}$ and $\theta_{13}$. For a given set of $\nu_{e}$ appearance data,
if the input value of $\theta_{23}$ is increased then the output
value of $\theta_{13}$ will be decreased.

Generalizing the definition of $\theta_{13}^{{\rm eff}}$ by including
higher order terms will be experiment-dependent since the NLO and NNLO terms have different
energy dependence, which means the spectrum of the neutrino beam and the efficiency
of neutrino detection have to be involved. However, by taking the approximation that the measurements are mainly determined by the total number of events\footnote{We have studied the validity of this approximation. For a T2K-like experiment, this approximation can keep valid when $N_{{\rm tot}}<{\cal O}(10^{3})$ or $\Delta E_{\nu}/E_{\nu}>{\cal O}(1\%)$---see more details in the appendix. }, we can integrate out the energy dependence at  NLO and NNLO, and get the following result:
\begin{equation}
\frac{1}{2}\sin^{2}2\theta_{13}^{{\rm eff}}\equiv s_{23}^{2}\sin^{2}2\theta_{13}-8\alpha J_{CP}f_{s}+8\alpha\frac{J_{CP}}{\tan\delta_{CP}}f_{c}+\alpha^{2}\sin^{2}2\theta_{12}c_{23}^{2}f_{2}. \label{eq:eff-8}
\end{equation}
Here $f_{s}$, $f_{c}$ and $f_{2}$ are numerical factors depending
on the experimental configurations, including baseline, neutrino beam,
detector, etc. For a T2K-like experiment, we have evaluated these $f$-factors, as listed in Tab.~\ref{tab:fff}. Note that the $f$-factors also depend on the mass ordering (normal/inverted), and the oscillation channels ($\nu_{\mu}\rightarrow\nu_{e}$  or $\overline{\nu}_{\mu}\rightarrow\overline{\nu}_{e}$).

\begin{table}[H]
\centering

\caption{\label{tab:fff}The $f$ factors in T2K.}

\begin{tabular}{ccccc}
\hline
 & \hspace{0.3cm}$\nu_{e}$ normal\hspace{0.3cm} & \hspace{0.3cm}$\nu_{e}$ inverted\hspace{0.3cm} & \hspace{0.3cm}$\overline{\nu}_{e}$ normal\hspace{0.3cm} & \hspace{0.3cm}$\overline{\nu}_{e}$ inverted\hspace{0.3cm}\tabularnewline
\hline
$f_{s}$ & 1.42083 & -1.62128 & -1.52638 & 1.47037\tabularnewline
$f_{c}$ & 0.0314521 & -0.0159949 & 0.107811 & -0.0327648\tabularnewline
$f_{2}$ & 2.98237 & 4.04585 & 3.51664 & 3.30968\tabularnewline
\hline
\end{tabular}
\end{table}
Eq.~(\ref{eq:eff-8}) can be used to approximately describe the
  correlations among the PMNS parameters $\theta_{13}$, $\theta_{23}$
  and $\delta_{CP}$ with very good accuracy. In what follows we discuss some important parameter correlations based on Eq.~(\ref{eq:eff-8}).
At  LO, only $\theta_{23}$ and $\theta_{13}$ are correlated, which has been discussed previously. Entering the NLO, the most important observable would be $\delta_{CP}$.
Here we would like to draw the reader's attention to the fact that $f_c$ presented here is very small\footnote{Although this is experiment dependent, it is actually a general feature in current  accelerator neutrino experiments.
The reason is due to the mismatch of the cosine oscillation mode
($\cos \Delta$) in the CP even term with the sine mode in Eq.\ (\ref{eq:eff}). A
more detailed explanation requires a closer look at the shape of the
neutrino flux, which is postponed to the appendix.}
compared to other $f$-factors.
As a consequence of small $f_c$,  the CP-even contribution (defined by the events generated by the oscillation term proportional to $\cos\delta_{CP}$) is small, which makes the measurement of $\delta_{CP}$ in the experiments such as T2K and DUNE actually sensitive to $\sin\delta_{CP}$ rather than $\cos\delta_{CP}$. This explains why the uncertainties of $\delta_{CP}$ in future measurements will be maximal or minimal if the true value of $\delta_{CP}$ is $\pm90^{\circ}$ or $0$, respectively---see,
e.g., Refs.~\cite{Nath:2015kjg,Ballett:2016daj,Fukasawa:2016yue,Coloma:2012wq}.

\begin{figure}
\centering

\includegraphics[width=11cm]{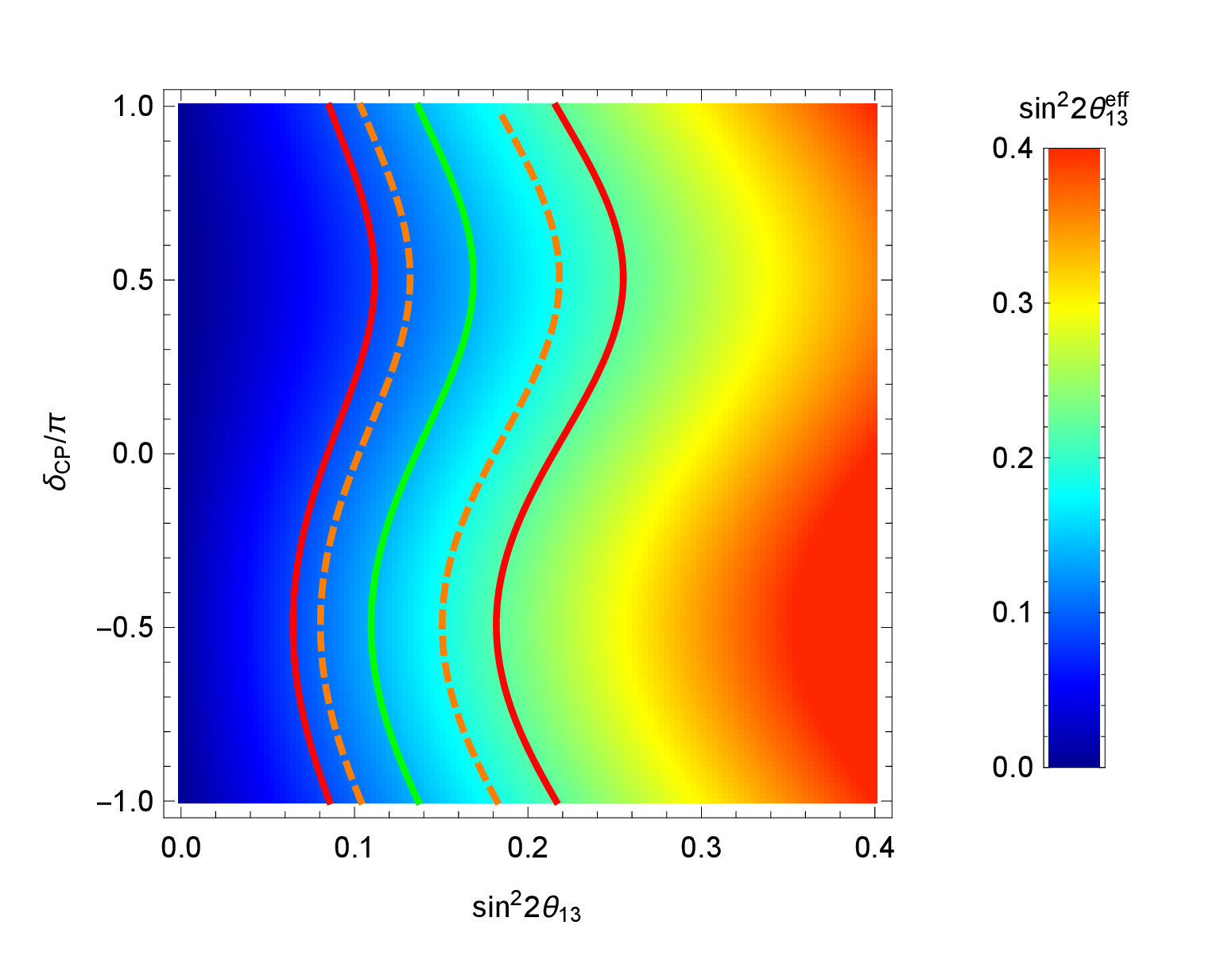}

\caption{Contour plot of $\sin^{2}2\theta_{13}^{{\rm eff}}(\theta_{13},\thinspace\delta_{CP})$.
% for the normal mass ordering.
The five contours from the left to the right correspond to $\sin^{2}2\theta_{13}^{{\rm eff}}=$ 0.087,
0.106, 0.139, 0.184 and 0.218, respectively. Normal mass ordering is assumed, with $(\theta_{12},\,\theta_{23},\,\alpha)=(34^{\circ},\,45^{\circ},\,0.031)$.}
\label{fig:Contour-plot}
\end{figure}

Next, let us look into the correlation of $\delta_{CP}$ and $\theta_{13}$.
Focusing on the $\nu_{\mu}\rightarrow\nu_{e}$ mode, we draw a contour plot of $\sin^{2}2\theta_{13}^{{\rm eff}}(\theta_{13},\thinspace\delta_{CP})$,
as shown in Fig.\ \ref{fig:Contour-plot}. Here we fix other parameters ($\theta_{23}$, $\theta_{12}$ and $\alpha$) at their best-fit values of the global fit \cite{Esteban:2016qun} and assume the normal ordering.

We choose five different
values for $\sin^{2}2\theta_{13}^{{\rm eff}}$ and plot the corresponding
contours. As one can see from these contours, they all appear in the
shape of sine curves, which originates from the $CP$-odd term (the
second term) in Eq.~(\ref{eq:eff-8}) that contains $\sin\delta_{CP}$.
The $CP$-even term can be neglected because $f_{c}^{{\rm T2K}}$
is very small.

Fig.\ \ref{fig:Contour-plot} implies that for a fixed value of $\theta_{13}^{{\rm eff}}$,
$\theta_{13}$ approximately increases with $\sin\delta_{CP}$; it
reaches the maximum (minimum) when $\delta_{CP}$ is close to $\pi/2$
($-\pi/2$).
%As a consequence, if $\theta_{13}$ measured in T2K is
%larger\footnote{Since the T2K value of $\theta_{13}$ is %$CP$-dependent, here ``larger''
%means for any value of $\delta_{CP}$ it is always larger than the
%reactor value. } than the measured reactor neutrino value of %$\theta_{13}$, the favored
%value of $\delta_{CP}$ in the
%combined analysis will necessarily be about $-\pi/2$.
%, which has been
%studied in the literature as maximal $CP$ violation \cite{He:2015xha,Joshipura:2015dsa}.
The shape of the contours resembles the constraint on $(\theta_{13},\thinspace\delta_{CP})$
published by the T2K collaboration (cf.\ Fig.~5 in Ref. \cite{T2K2014PRL},
Fig.~39 in Ref. \cite{Abe:2017vif}, or Fig.~\ref{fig:fit-t2k}
in this paper).
Indeed, given the observed numbers of events (32 $\nu_e$ and 4 $\overline{\nu}_e$) in T2K,
the constraint on $(\theta_{13},\thinspace\delta_{CP})$ can be readily
 reproduced merely from a fit on  $\sin^{2}2\theta_{13}^{{\rm eff}}$, which we present in the appendix.
 %, given the T2K constraint on .

%\subsection{\label{sec:41}Analytical discussion}
Another interesting topic is about the maximal CP phase ($\delta_{CP}=-\pi/2$) which has been the best fit value when combining the T2K result and the reactor measurements since 2013 \cite{T2K2014PRL}.
Based on Eq.~(\ref{eq:eff-8}),
we can argue analytically on when the maximal CP phase would appear. Since the $CP$-even
term and the $\alpha^{2}$ term are much smaller than the other terms,
we neglect them in the following discussion:
\begin{equation}
\frac{1}{2}S^{2}\equiv\frac{1}{2}\sin^{2}2\theta_{13}^{{\rm eff}}\approx s_{23}^{2}\sin^{2}2\theta_{13}-8\alpha J_{CP}f_{s}.\label{eq:eff-29}
\end{equation}
For fixed values of $\theta_{13}$, $\theta_{23}$ and $\theta_{12}$,
$S^{2}$ has a maximum $S_{{\rm max}}^{2}$ and a minimum $S_{{\rm min}}^{2}$
at $\delta_{CP}=-\pi/2$ and $\pi/2$, respectively. If the measured
value of $S^{2}$ falls into the range of $(S_{{\rm min}}^{2},\ S_{{\rm max}}^{2})$,
then one can compute the corresponding value of $\sin\delta_{CP}$
by
\begin{equation}
\sin\delta_{CP}=\frac{2\sin^{2}2\theta_{13}s_{23}^{2}-S^{2}}{2\alpha f_{s}c_{13}\sin2\theta_{12}\sin2\theta_{13}\sin2\theta_{23}}.\label{eq:eff-30}
\end{equation}
If the measured $S^{2}$ is not in the range $(S_{{\rm min}}^{2},\ S_{{\rm max}}^{2})$,
then there may be a tension between the
$\nu_{e}$ ($\overline{\nu}_{e}$) appearance and disappearance
data. In this case, if these data are combined, the best-fit value of $\delta_{CP}$
will be pushed to $\pm\pi/2$.
%, which
%has some important implications on neutrino flavor symmetries and
%models
%\cite{Harrison:2002et,Ma:2002ce,Babu:2002dz,Ma:2002ge,Grimus:2003yn,Nishi:2013jqa, Ma:2013mga, Fraser:2014yha,He:2015gba,Ma:2015gka,DiIura:2015kfa,Mohapatra:2015gwa, Zhou:2014sya,Joshipura:2015dsa,He:2015xha, Zhao:2017yvw,Nishi:2016wki,Chen:2015siy, Fukuyama:2017qxb,He:2012yt,Rodejohann:2017lre,Rodejohann:2015hka}.
Neglecting possible tensions, from the measurement of $S^{2}$ in
$\nu_{e}$ ($\overline{\nu}_{e}$)
appearance experiments one can obtain a $CP$-dependent measurement
of $\theta_{13}$, which is
\begin{equation}
\sin2\theta_{13}\approx\frac{\beta\sin\delta_{CP}+\sqrt{2s_{23}^{2}S^{2}+\beta^{2}\sin^{2}\delta_{CP}}}{2s_{23}^{2}}\label{eq:eff-32}, \mbox{ where }
%\end{equation}
%where
%\begin{equation}
\beta\equiv\alpha f_{s}c_{13}\sin2\theta_{12}\sin2\theta_{23}.%\label{eq:eff-38}
\end{equation}
Although the measurement is $CP$-dependent, since $-1\leq\sin\delta_{CP}\leq1$,
$\theta_{13}$ measured in $\nu_{e}$ ($\overline{\nu}_{e}$) appearance
experiments should be in the following range:
\begin{equation}
\frac{\sqrt{2s_{23}^{2}S^{2}+\beta^{2}}-|\beta|}{2s_{23}^{2}}\apprle\theta_{13}\apprle\frac{\sqrt{2s_{23}^{2}S^{2}+\beta^{2}}+
|\beta|}{2s_{23}^{2}}.
\label{eq:eff-34}
\end{equation}
Again, if the measurement of $\theta_{13}$ from reactor neutrinos is lower
(or higher) than the lower (or upper) bound in Eq.~(\ref{eq:eff-34}),
then the combined data fitting always prefers maximal CP violating
values of $\delta_{CP}$.

In summary, the effective $\theta_{13}$ analytically shows  correlations of $\delta_{CP}$, $\theta_{13}$ and $\theta_{23}$ in the reactor and accelerator neutrino experiments. According to the dependence of $\theta^{\rm eff}_{13}$ on $\delta_{CP}$, $\theta_{13}$ and $\theta_{23}$, one can understand why the CP-even oscillation term contributes to the $\nu_e$ appearance data much less than the CP-odd term. This implies the uncertainties of
$\delta_{CP}$ measurements will be maximal or minimal if the true value of $\delta_{CP}$ is $\pm90^{\circ}$ or 0, respectively. Moreover, using the effective $\theta_{13}$, we can derive the upper and lower bounds (\ref{eq:eff-34}) of reactor $\theta_{13}$, beyond which the combination of reactor and accelerator neutrino data should prefer maximal CP violation. This can be verified in the numerical analyses in the next section.

\section{Numerical Results\label{sec:applications}}

In this section, we adopt the numerical approach to study the phenomenologies regarding the determination and correlation of
$\delta_{CP}$, $\theta_{13}$, $\theta_{23}$ and the mass ordering \cite{Minakata:2002jv,Huber:2003pm,
Minakata:2003wq,Mena:2004sa,Ghosh:2012px,Choubey:2013xqa}.

Since $\theta_{13}$ and $\theta_{23}$ can be measured independently in (anti)neutrino disappearance experiments and the measurements will be frequently used in this section, we list the recent results in Tab.~\ref{tab:theta1323}.  The NO$\nu$A measurement \cite{Adamson:2017qqn} of $\theta_{23}$ contains two best-fit solutions. Henceforth we will refer to the two solutions in the $\theta_{23}<45^{\circ}$ and $\theta_{23}>45^{\circ}$ octants as NO$\nu$A$^-$ and NO$\nu$A$^+$, respectively\footnote{With the leading term
in the muon-neutrino survival probability proportional to $\sin^2 2\theta_{23}$, solutions in
both octants are naturally expected.}.
As future data will pin down the true values, it is of interest here to analyze the impact of the
possible true values on the current and possible future hints of mass ordering and CP phase.

\begin{table}[h]
\centering

\caption{\label{tab:theta1323}Recent measurements of $\theta_{13}$ and $\theta_{23}$. N and I stand for the normal and inverted mass ordering respectively.}

\begin{tabular*}{16cm}{@{\extracolsep{\fill}}rccc}
\hline
\hline
 & Daya Bay \cite{An:2016ses} & \hspace{0.3cm}RENO \cite{Seo:2016dbz}\hspace{0.3cm} & \hspace{0.3cm}Double Chooz \cite{Matsubara:2016prb}\hspace{0.3cm}\tabularnewline
\hline
$\sin^{2}2\theta_{13}$ & $0.0841\pm0.0033$ & $0.088\pm0.011$ & $0.111\pm0.018$\tabularnewline
$\theta_{13}/^{\circ}$ & $8.43_{-0.17}^{+0.17}$ & $8.62{}_{-0.57}^{+0.54}$ & $9.73_{-0.85}^{+0.79}$\tabularnewline
\hline\hline
 & \hspace{0.2cm}T2K \cite{Abe:2017bay} & NO$\nu$A$^{-}$  \cite{Adamson:2017qqn}& NO$\nu$A$^{+}$  \cite{Adamson:2017qqn} \tabularnewline\hline
$\sin^{2}\theta_{23}$ (N) & $0.532_{-0.068}^{+0.046}$ & $0.404_{-0.022}^{+0.030}$  &  $0.624_{-0.030}^{+0.022}$\tabularnewline
$\theta_{23}/^{\circ}$ (N) & $46.8_{-3.9}^{+2.7}$ & $39.5_{-1.3}^{+1.7}$  & $52.2_{-1.8}^{+1.3}$\tabularnewline
$\sin^{2}\theta_{23}$ (I) & $0.534_{-0.066}^{+0.043}$ & $0.398_{-0.022}^{+0.030}$  &  $0.618_{-0.030}^{+0.022}$\tabularnewline
$\theta_{23}/^{\circ}$ (I) & $46.9_{-3.8}^{+2.5}$ & $39.1_{-1.3}^{+1.8}$  &  $51.8_{-1.8}^{+1.3}$\tabularnewline
\hline
\hline
\end{tabular*}

\end{table}

For the (anti)neutrino appearance data fitting, we only focus on
  the T2K experiment, whose experimental parameters are more easy to
  access, and which has  larger event numbers.
Furthermore, the purpose of this paper
is to note general features of parameter correlations of the
parameters.
For analyses including all available neutrino data, we would like to refer to the global fit work \cite{Capozzi:2016rtj,Esteban:2016qun,deSalas:2017kay}.

We use the appearance data published in \cite{Abe:2017uxa} (taken from its Fig.~3) to construct the $\chi^2$-function of $(\theta_{23},\,\theta_{13},\,\delta_{CP})$ while the other parameters not of interest are simply fixed at the best-fit values in the global fit. The details of event rate computation (including the cross sections, the neutrino beams) are covered by the appendix.
The $\chi^2$-function computed from the T2K data will be referred to  as  $\chi^2_{\rm{T2K}}$ below. With this $\chi^2$ function, we proceed to the following phenomenological studies.

\subsection{Maximal CP violation}

As we have concluded in the analytical discussion, the appearance of maximal CP violation depends on the status of the direct measurement of $\theta_{13}$.  If $\theta_{13}$ measured is out of the bounds in Eq.~(\ref{eq:eff-34}), then the best fit of $\delta_{CP}$ stays at maximal values.
\begin{figure}
\centering

\includegraphics[width=10cm]{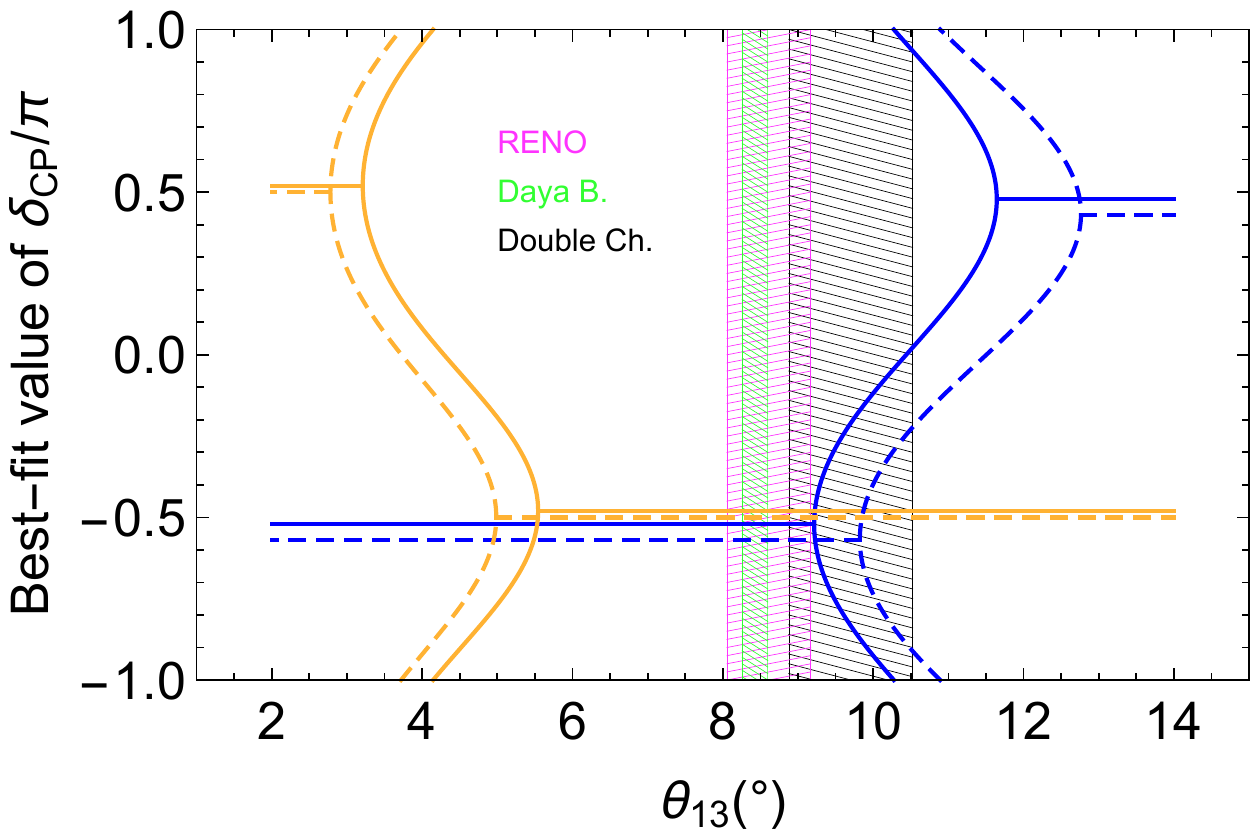}

\caption{The best-fit value of $\delta_{CP}$ from T2K $\nu_{e}$ (blue curve)
and $\overline{\nu}_{e}$ (orange curve) data \cite{Abe:2017vif} as a
function of $\theta_{13}$. Solid/dashed curves
are for normal/inverted mass ordering. For $5.5^{\circ}<\theta_{13}<9.2r^{\circ}$,
both $\nu_{e}$ and $\overline{\nu}_{e}$ data prefer a maximal CP
violation value ($-\pi/2$) of $\delta_{CP}$. \label{fig:whenMCP}}
\end{figure}
In Fig.\ \ref{fig:whenMCP} we show the best-fit value of $\delta_{CP}$ in the T2K $\nu_{e}$ ($\overline{\nu}_{e}$) appearance data, which
confirms the analytical arguments. The value of $\delta_{CP}$
is computed by
\begin{equation}
\min_{\delta_{CP}}\chi_{{\rm T2K}}^{2}(\delta_{CP},\ \theta_{13},\ \theta_{23})\,,
\label{eq:eff-31}
\end{equation}
where $\theta_{13}$ varies from $2^{\circ}$ to $14^{\circ}$, and $\theta_{23}$ is fixed at $46.6^{\circ}$. We compute the best-fit value of $\delta_{CP}$ for both neutrino
and antineutrino data, normal and inverted mass ordering, plotted
by blue and orange, solid and dashed curves in Fig.\ \ref{fig:whenMCP}
respectively.
r

Note that in the T2K data, the observed number of 32 $\nu_e$ (or 4 $\overline{\nu}_e$)
events is larger (or smaller) than the expected number, which should
be 28.7 (6.0), 24.2 (6.9), or 19.6 (7.7) for $\delta_{CP}=$ $-\pi/2$,
$0$, or $\pi/2$, respectively \cite{Magaletti:2017ltp}. Since the expected number of $\nu_e$ ($\overline{\nu}_e$) decreases (or increases) with $\sin\delta_{CP}$, both the excess of observed $\nu_e$ and the deficit of $\overline{\nu}_e$ favor minimal $\sin\delta_{CP}$, i.e.\ $\delta_{CP}=-\pi/2$.

rAs shown in the plot, the T2K $\nu_{e}$ data favors
a range of $\theta_{13}$ in $[9.2^{\circ},\thinspace11.6^{\circ}]$
for the normal ordering or $[9.8^{\circ},\thinspace12.7^{\circ}]$
for the inverted ordering. Currently reactor neutrino experiments
including Daya Bay \cite{An:2016ses}, RENO \cite{Seo:2016dbz} and
Double Chooz \cite{Matsubara:2016prb} all have measured smaller values
of $\theta_{13}$, marked by green, magenta and black bands (1$\sigma$
CL) in Fig.\ \ref{fig:whenMCP}. Therefore when the T2K $\nu_{e}$
data is combined with any of the reactor measurements, the best-fit
value of $\delta_{CP}$ will necessarily become about $-\pi/2$. For
the T2K $\overline{\nu}_{e}$ data, the favored range is $[3.2^{\circ},\thinspace5.5^{\circ}]$
(normal) or $[2.8^{\circ},\thinspace5.0^{\circ}]$ (inverted), which
is smaller than the reactor measurements. So combining the T2K $\overline{\nu}_{e}$
data with any of the reactor measurements also leads to about $-\pi/2$
for the best-fit value of $\delta_{CP}$. Note that due to corrections
from the $CP$-even term, the actual value deviates from $-\pi/2$
by about ${\cal O}(0.05\pi)$.

\begin{figure}[h]
\centering

\includegraphics[width=7.5cm]{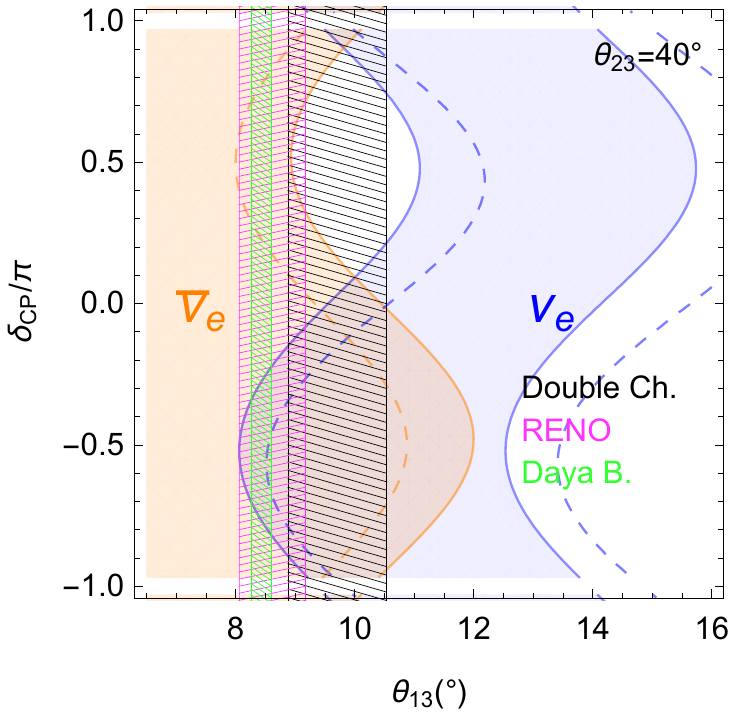}\ \includegraphics[width=7.5cm]{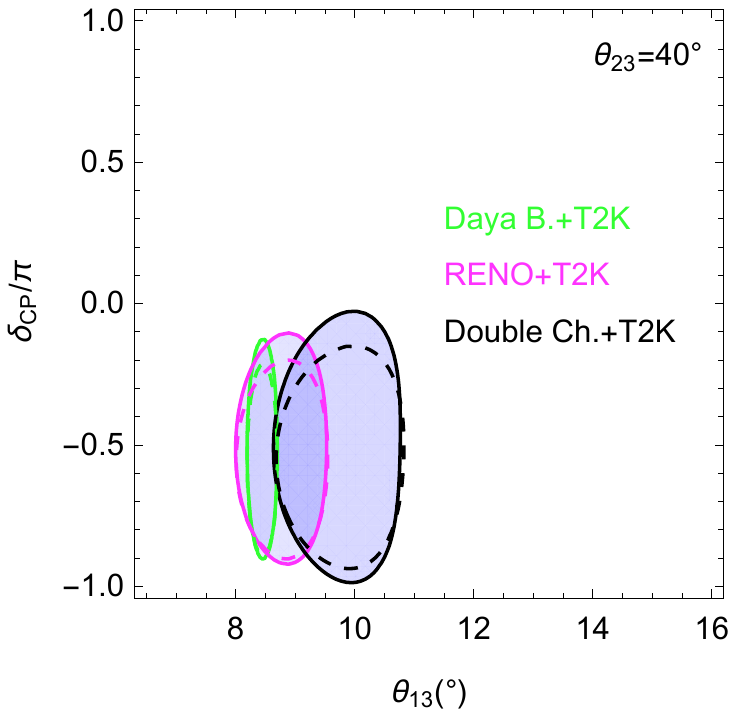}

\includegraphics[width=7.5cm]{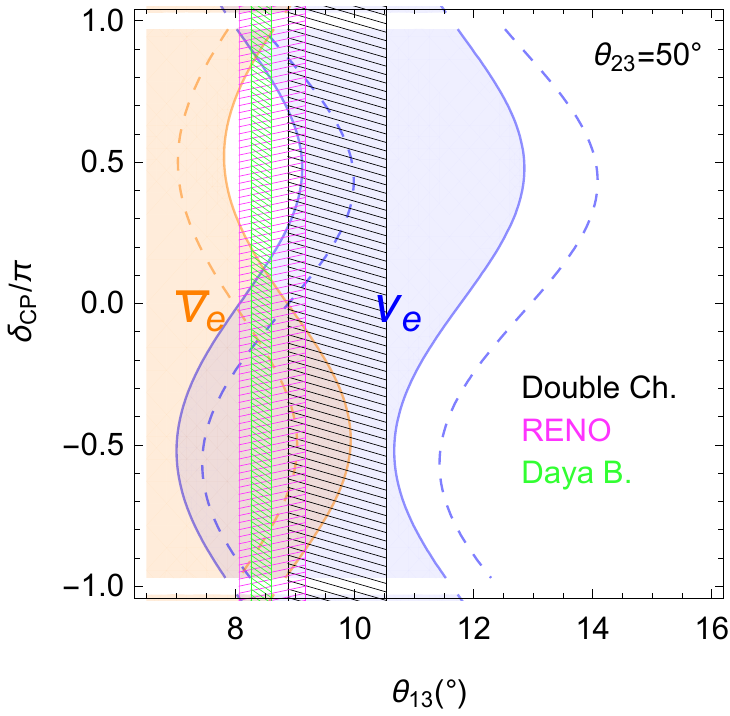}\ \includegraphics[width=7.5cm]{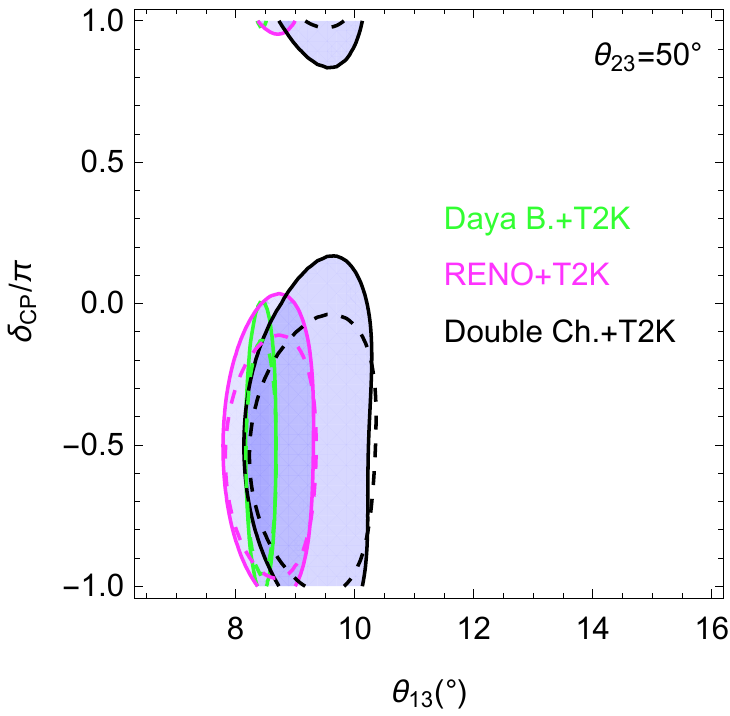}

\caption{{\it Left panel:}
constraints on $(\theta_{13},\thinspace\delta_{CP})$ from T2K \cite{Abe:2017vif}
and constraints on $\theta_{13}$ from reactor neutrino experiments.
 Solid/dashed curves are for normal/inverted mass ordering, blue/orange
curves for $\nu_{e}$/$\overline{\nu}_{e}$ appearance data, respectively. {\it Right panel:}
Fit results on $(\theta_{13},\thinspace\delta_{CP})$ when
T2K neutrino and antineutrino data are combined with reactor data.
The upper plots apply for the NO$\nu$A$^-$ solution of $\theta_{23}$, the lower plots for
NO$\nu$A$^+$.
 \label{fig:combine13}}
\end{figure}

\subsection{\label{sec:42}Correlation of $\theta_{13}$ and $\delta_{CP}$}
The two-parameter fit on $(\theta_{13},\thinspace\delta_{CP})$
 will show the correlation of these two parameters.
By fixing $\theta_{23}$ at certain values in $\chi_{{\rm T2K}}^{2}(\delta_{CP},\ \theta_{13},\ \theta_{23})$ and allowing $\Delta\chi^{2}\leq2.3$ in the fit,
we plot the 68\% CL constraints on $(\theta_{13},\thinspace\delta_{CP})$,
%from the T2K $\nu_{e}$ and $\overline{\nu}_{e}$ appearance data,
shown in the left panel of Fig.\ \ref{fig:combine13} by the blue (for $\nu_{e}$) and orange (for $\overline{\nu}_{e}$)
regions. The regions
within solid (dashed) lines assume normal (inverted) mass ordering.

As we can expect from Eq.~(\ref{eq:eff-29}), which shows the dependence
of $\sin^{2}2\theta_{13}^{{\rm eff}}$ on $\sin\delta_{CP}$, the
orange and blue bounds all have the shapes of sine curves. The curves
of $\nu_{e}$ and $\overline{\nu}_{e}$ are bent to opposite directions: the
$\nu_{e}$ curves have minimal $\theta_{13}$ at
$\delta_{CP}\approx-\pi/2$,
while for $\overline{\nu}_{e}$ it is maximal. As a result, the orange
region and the blue region have some overlap where $\delta_{CP}$
is mainly negative. This implies that combined fitting of $\nu_{e}$
and $\overline{\nu}_{e}$ appearance data favors negative $\delta_{CP}$.
In the plots we also show the 1$\sigma$ CL bounds on $\theta_{13}$ from
reactor neutrino experiments. When $\theta_{23}$ is fixed at $40^{\circ}$,
the bounds from Daya Bay \cite{An:2016ses} and RENO \cite{Seo:2016dbz}
are below the overlap of the T2K $\nu_{e}$ and $\overline{\nu}_{e}$
constraints while the Double Chooz bound \cite{Matsubara:2016prb}
is well compatible with the $\nu_{e}$-$\overline{\nu}_{e}$ overlap.
However, one should notice that this depends on the value of $\theta_{23}$.
If $\theta_{23}$ is fixed at the NO$\nu$A$^+$ value, as shown in the lower
plots in Fig.\ \ref{fig:combine13}, the $\nu_{e}$-$\overline{\nu}_{e}$
overlap covers all the reactor bounds.

In the right panel of Fig.\ \ref{fig:combine13}, we combine the
T2K data with reactor neutrino data by
\begin{equation}
\chi^{2}(\theta_{13},\thinspace\delta_{CP})=\chi_{{\rm reactor}}^{2}(\theta_{13})+\chi_{{\rm T2K},\ \nu_{e}}^{2}(\theta_{13},\thinspace\delta_{CP})+\chi_{{\rm T2K},\ \overline{\nu}_{e}}^{2}(\theta_{13},\thinspace\delta_{CP})
,\label{eq:eff-16}
\end{equation}
where the first, second and last terms are constraints from reactor
neutrino data, T2K $\nu_{e}$ and $\overline{\nu}_{e}$ data, respectively.
The $\chi^{2}$-function of $\theta_{13}$
from reactor neutrino data we adopt is
\begin{equation}
\chi_{{\rm reactor}}^{2}(\theta_{13})=\left(\frac{\sin^{2}2\theta_{13}-\sin^{2}2\theta_{13}^{0}}{\sigma_{13}}\right)^{2}.\label{eq:eff-26}
\end{equation}
Note that Eq.~(\ref{eq:eff-26}) assumes the distribution of $\sin^{2}2\theta_{13}$
to be Gaussian
%\footnote{Actually with limited event numbers, neither $\sin^{2}2\theta_{13}$
%nor $\theta_{13}$ measured in reactor neutrino experiments is Gaussian.
%When the statistical and systematic uncertainties are small enough,
%both are approximately in Gaussian distributions but for $\sin^{2}2\theta_{13}$
%the approximation is better.},
with a central value $\sin^{2}2\theta_{13}^{0}$ and the standard
deviation $\sigma_{13}$ given in Tab.~\ref{tab:theta1323}.

\begin{comment}
\begin{table}[h]
\centering

\caption{\label{tab:theta13}Measurements of $\theta_{13}$ from reactor neutrino
experiments.}

\begin{tabular}{cccc}
\hline
 & \hspace{0.3cm}Daya Bay \cite{An:2016ses}\hspace{0.3cm} & \hspace{0.3cm}RENO \cite{Seo:2016dbz}\hspace{0.3cm} & \hspace{0.3cm}Double Chooz \cite{Matsubara:2016prb}\hspace{0.3cm}\tabularnewline
\hline
$\sin^{2}2\theta_{13}$ & $0.0841\pm0.0033$ & $0.088\pm0.011$ & $0.111\pm0.018$\tabularnewline
$\theta_{13}/^{\circ}$ & $8.43_{-0.17}^{+0.17}$ & $8.62{}_{-0.57}^{+0.54}$ & $9.73_{-0.85}^{+0.79}$\tabularnewline
\hline
\end{tabular}

\end{table}
\end{comment}

For each reactor neutrino experiment listed in Tab.~\ref{tab:theta1323},
we perform a $\chi^{2}$-fit combined with the T2K data and compute
the corresponding 68\% CL constraints on $(\theta_{13},\thinspace\delta_{CP})$,
presented in the right panel of Fig.\ \ref{fig:combine13} by the green,
magenta, and black contours (dashed for inverted mass ordering) for
Daya Bay, RENO and Double Chooz, respectively. For all three reactor
neutrino experiments the results favor negative $\delta_{CP}$. Note
that the significance depends on $\theta_{23}$: when $\theta_{23}$
increases, the bounds on $\delta_{CP}$ expand and the significance
of CP violation decreases.
One finds from the plot that the influence of the true value of $\theta_{13}$
is not dramatic but still noteworthy; for $\theta_{23}$ in the second octant (NO$\nu$A$^+$)
the effect is
however slightly larger (note that in Fig.\ \ref{fig:combine13} the overlap regions
of neutrino and antineutrino appearance data fits in this case better
with the reactor determinations of $\theta_{13}$). The origin of this
behavior is easy to identify in Eq.\ (\ref{eq:eff-3}), where $\sin^2
\theta_{23}$ appears.
Hence, a value of $\theta_{23}$ in the upper octant
may thus reduce the current significance of the hints for maximal CP
violation.

\begin{figure}
\centering

\includegraphics[width=7.5cm]{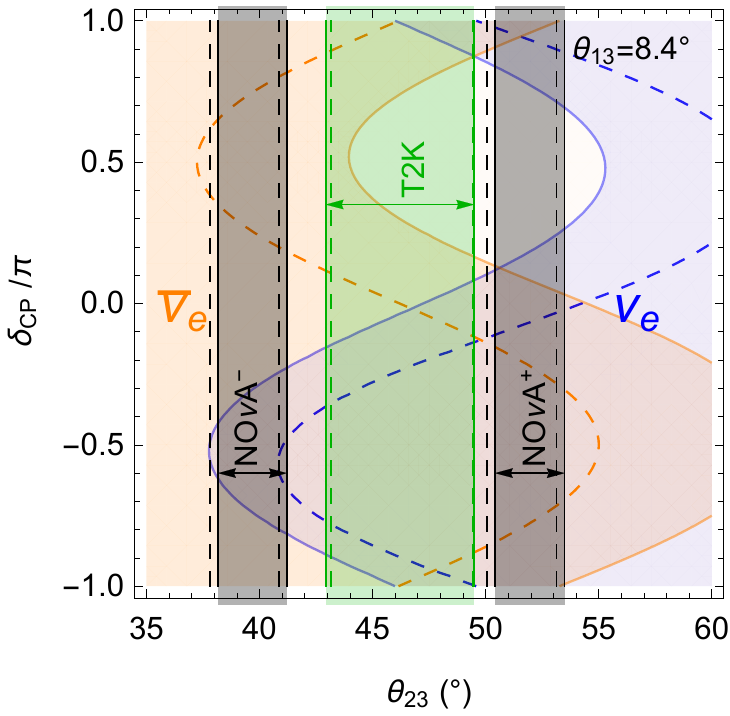}\ \includegraphics[width=7.5cm]{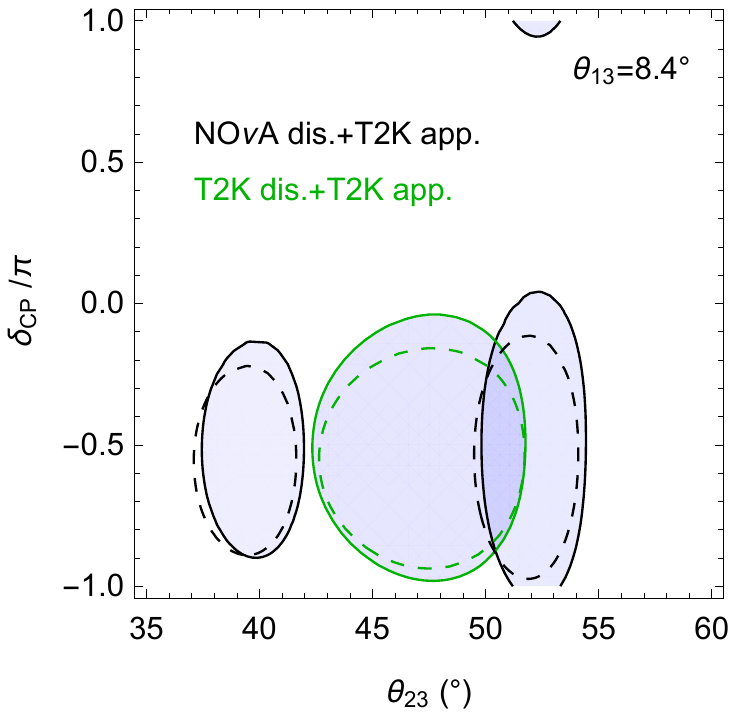}

\includegraphics[width=7.5cm]{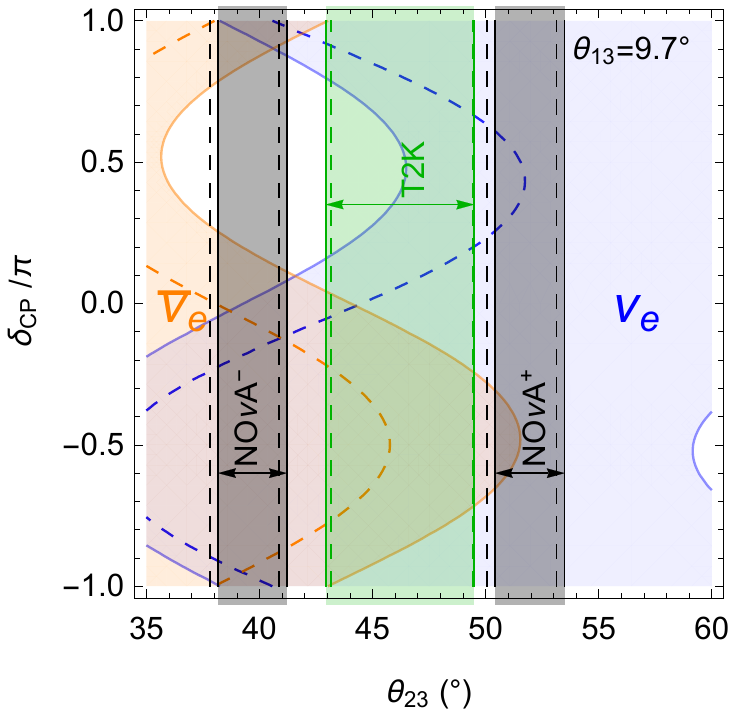}\ \includegraphics[width=7.5cm]{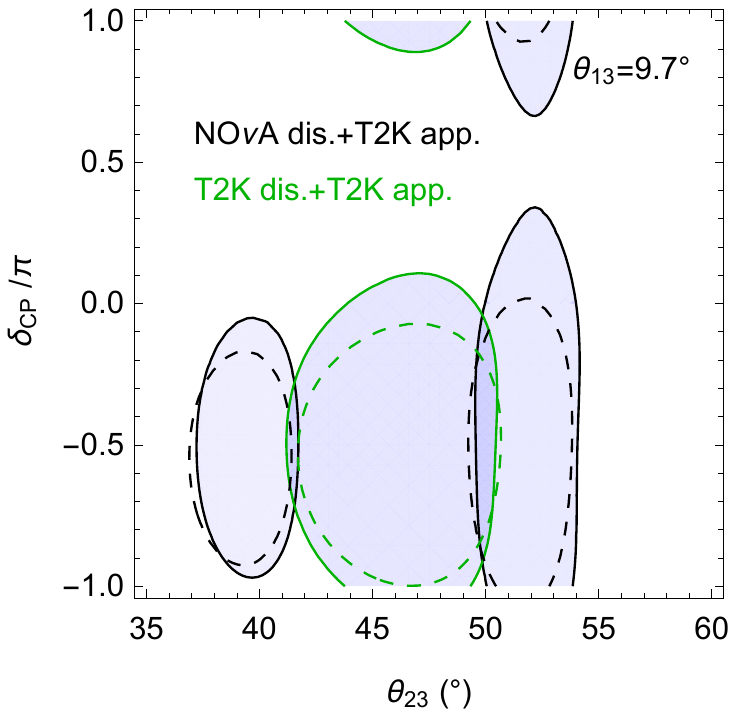}

\caption{Constraints on $(\theta_{23},\thinspace\delta_{CP})$ from T2K appearance
data combined with $\nu_{\mu}$ disappearance data from T2K and NO$\nu$A.
Solid/dashed curves are for normal/inverted mass ordering, blue/orange
curves for $\nu_{e}$/$\overline{\nu}_{e}$ appearance data, respectively.
The upper plots are for the best-fit value of $\theta_{13}$ from Daya Bay/RENO, the
lower ones for the best-fit value of $\theta_{13}$ from Double Chooz.
\label{fig:combine23}}
\end{figure}

\subsection{\label{sec:43}Correlation of $\theta_{23}$ and $\delta_{CP}$}
As we have seen, $\theta_{23}$ also plays an important role in the
measurement of $\delta_{CP}$ in accelerator neutrino experiments.
Thus we shall study the relation of $\delta_{CP}$ and $\theta_{23}$ in
more detail. Similar to Eq.~(\ref{eq:eff-16}),
we have the following combined $\chi^{2}$-function:
\begin{equation}
\chi^{2}(\theta_{23},\thinspace\delta_{CP})=\chi_{{\rm Acc.\thinspace Dis.}}^{2}(\theta_{23})+\chi_{{\rm T2K},\ \nu_{e}}^{2}(\theta_{23},\thinspace\delta_{CP})+\chi_{{\rm T2K},\ \overline{\nu}_{e}}^{2}(\theta_{23},\thinspace\delta_{CP})
.\label{eq:eff-16-1}
\end{equation}
Here $\chi_{{\rm Acc.\thinspace Dis.}}^{2}$ denotes the $\chi^{2}$-function
of $\theta_{23}$ constrained by accelerator neutrino disappearance
experiments. In Tab.~\ref{tab:theta1323}, we list the measurements
of $\theta_{23}$ from T2K and NO$\nu$A. For simplicity, in $\chi_{{\rm Acc.\thinspace Dis.}}^{2}$
we assume a Gaussian distribution of $\sin^{2}\theta_{23}$,
\begin{equation}
\chi_{{\rm Acc.\thinspace Dis.}}^{2}(\theta_{23})=\left(\frac{\sin^{2}\theta_{23}-\sin^{2}\theta_{23}^{0}}{\sigma_{23}}\right)^{2},\label{eq:eff-36}
\end{equation}
where $\sigma_{23}$ is taken as the mean value of the upper and lower uncertainties.
\begin{comment}
\begin{table}[H]
\centering

\caption{\label{tab:theta23}Measurements of $\theta_{23}$ from accelerator
neutrino disappearance experiments.}

\begin{tabular}{ccccc}
\hline
 & \hspace{0.2cm}T2K (normal)\hspace{0.2cm} & \hspace{0.2cm}T2K (inverted)\hspace{0.2cm} & \hspace{0.4cm}NO$\nu$A (normal)\hspace{0.4cm} & \hspace{0.4cm}NO$\nu$A (inverted)\hspace{0.4cm}\tabularnewline
\hline
$\sin^{2}\theta_{23}$ & $0.532_{-0.068}^{+0.046}$ & $0.534_{-0.066}^{+0.043}$ & $0.404_{-0.022}^{+0.030}$ or $0.624_{-0.030}^{+0.022}$\  & \ $0.398_{-0.022}^{+0.030}$ or $0.618_{-0.030}^{+0.022}$\tabularnewline
$\theta_{23}/^{\circ}$ & $46.8_{-3.9}^{+2.7}$ & $46.9_{-3.8}^{+2.5}$ & $39.5_{-1.3}^{+1.7}$ or $52.2_{-1.8}^{+1.3}$ & $39.1_{-1.3}^{+1.8}$ or $51.8_{-1.8}^{+1.3}$\tabularnewline
\hline
\end{tabular}

\end{table}
\end{comment}
In the left panel of Fig.\ \ref{fig:combine23}, we show separate
constraints on $(\delta_{CP},\thinspace\theta_{23})$ from the three
terms in Eq.~(\ref{eq:eff-16-1}). The constraints from the T2K $\nu_{e}$
and $\overline{\nu}_{e}$ appearance data are presented by blue and
orange regions. We can explain
this behavior by rewriting Eq.~(\ref{eq:eff-29}) as
\begin{equation}
s_{23}^{2}\approx\frac{S^{2}}{2\sin^{2}2\theta_{13}}+\frac{\beta}{\sin2\theta_{13}}\sin\delta_{CP},\label{eq:eff-37}
\end{equation}
where $\beta$ has been defined in Eq.~(\ref{eq:eff-32}). Although
$\beta$ contains $\sin2\theta_{23}$, for $40^{\circ}<\theta_{23}<50^{\circ}$,
the dependence of $\beta$ on $\theta_{23}$ is very weak $(0.985\leq\sin2\theta_{23}\leq1)$
so one can approximately treat it as a constant with respect to $\theta_{23}$.
Therefore Eq.~(\ref{eq:eff-37}) shows that  the constraints on $s_{23}^{2}$
from the $\nu_{e}$ and $\overline{\nu}_{e}$ appearance data are
$\delta_{CP}$ dependent and the dependence is described by a NLO
correction proportional to $\sin\delta_{CP}$. Because $\beta$ in
Eq.~(\ref{eq:eff-37}) has opposite signs for $\nu_{e}$ and $\overline{\nu}_{e}$,
the curves are bent to opposite directions so that the blue and orange
regions have overlap around $\delta=-\pi/2$. The overlap can be
approximately regarded as the region of preferred values of $(\delta_{CP},\thinspace\theta_{23})$
by the T2K appearance data. Note that this region depends on the fixed
value of $\theta_{13}$ used in Eq.~(\ref{eq:eff-16-1}). We choose
two values $\theta_{13}=8.4^{\circ}$ and $\theta_{13}=9.7^{\circ}$
(which are the central values of the Daya Bay/RENO and Double Chooz measurements,
respectively) to draw the plots in Fig.\ \ref{fig:combine23}. For
both cases, $\theta_{23}$ measured from the T2K $\nu_{\mu}$ ($\overline{\nu}_{\mu}$)
disappearance experiment (the green band) is well compatible with
the appearance data. The NO$\nu$A disappearance data favor non-maximal
$\theta_{23}$, showing two separate 1$\sigma$ bands (black) on the
plots. For $\theta_{13}=8.4^{\circ}$, the band of NO$\nu$A$^+$
is well compatible with the T2K appearance data while the other band
NO$\nu$A$^-$ is not. However, for
$\theta_{13}=9.7^{\circ}$, the preference moves to the band in the
lower octant. In the right panel of Fig.\ \ref{fig:combine23} we
show 1$\sigma$ constraints from a combined fit. For the combination
of the T2K disappearance data with the appearance data, the 1$\sigma$
bound of $\delta_{CP}$ is $(-0.9\pi,\thinspace-0.1\pi)$ when $\theta_{13}$
is fixed at $8.4^{\circ}$ and it expands to $(-1.1\pi,\thinspace0.1\pi)$
if $\theta_{13}$ is fixed at $9.7^{\circ}$. The combination of the
NO$\nu$A disappearance data with T2K appearance data, however, is
more sensitive to the value of $\theta_{13}$. When $\theta_{13}$
is $8.4^{\circ}$, the two separate regions in black contours have
$\delta_{CP}$ limited in $(-0.9\pi,\thinspace-0.1\pi)$ or $(-1.0\pi,\thinspace0\pi)$.
When $\theta_{13}$ is increased to $9.7^{\circ}$, both
expand by about $0.1\pi$ or $0.2\pi$.

One finds from the plot that the influence of the true value of $\theta_{13}$
is not dramatic but nevertheless noteworthy; if $\theta_{13}$ takes the large value,
the effect of $\theta_{23}$ from NO$\nu$A$^+$ is however slightly larger
%if $\theta_{23}$ lies in the second octant and if in
%addition $\theta_{13}$ is small, the effect is
%however slightly larger
(note that in Fig.\ \ref{fig:combine23} the overlap regions
of neutrino and antineutrino appearance data fit in this case show a slight
tension with the NO$\nu$A$^+$ solution).
Hence, if $\theta_{23}$ lies in the second octant and $\theta_{13}$ is
large (i.e.\ given by value of Double Chooz), then the significance  of the hints for maximal CP
violation is reduced.

\subsection{\label{sec:CPVfuture}Discovery of CP violation in the near future}

The current $\nu_{e}$ and $\bar{\nu}_{e}$ data of T2K, which was
collected from 2010 to 2016 with $1.5\times 10^{21}$ protons-on-target
(POT), shows only a moderate hint for CP Violation (CPV). Here we will
investigate how this might increase with future data (see also \cite{Ghosh:2014zea,Ghosh:2015tan}).
The T2K experiment will keep on collecting data until 2026 and will
obtain more than 10 times data ($20\times 10^{21}$ POT)
\cite{Abe:2016tii}, before the next generation of accelerator
neutrino experiments (e.g.\ DUNE \cite{Acciarri:2015uup}, T2HK \cite{Abe:2014oxa}
or T2HKK \cite{Abe:2016ero}) takes over.

To study this issue, we use the following $\chi^2$-function to compute
the pre-DUNE significance of CPV in the data:
\begin{equation}
\chi_{{\rm CPV}}^{2}\left(\theta_{23},\thinspace\theta_{13}\right)=\min_{\delta_{CP}=0\thinspace{\rm or}\thinspace\pi}\left[\chi_{{\rm \nu_{e}+\overline{\nu}_{e}}}^{2}\left(\theta_{23},\thinspace\theta_{13},\thinspace\delta_{CP}\right)\right]-\min_{\delta_{CP}\in[-\pi,\pi]}\left[\chi_{{\rm \nu_{e}+\overline{\nu}_{e}}}^{2}\left(\theta_{23},\thinspace\theta_{13},\thinspace\delta_{CP}\right)\right],
\label{eq:eff-0916}
\end{equation}
where $\chi_{{\rm \nu_{e}+\overline{\nu}_{e}}}^{2}$ is the sum of the
$\chi^2$-functions of the $\nu_e$ and $\overline{\nu}_{e}$ data. Our
definition $\chi_{{\rm CPV}}^{2}$ is the difference of how good a fit
with CP conservation is with respect to the CP violating best-fit
value.
For
the future data of T2K collected until 2026, we take the estimated
numbers from Ref.~\cite{Abe:2016tii} for the normal mass ordering,
including 558.7 $\nu_e$ events and 115.8 $\overline{\nu}_{e}$ events,
with a background  of 110.1 ($\nu_e$)+63.5 ($\overline{\nu}_{e}$)
events. These numbers are assuming maximal CP violation
($\delta_{CP}=-\pi/2$) and equal exposure in the neutrino and
antineutrino modes, i.e.\ $1.0\times10^{22}$ POT for $\nu_e$ and
$1.0\times10^{22}$ for $\overline{\nu}_{e}$. We will refer to this
data as T2K 2026 $(1:1)$. It may be favorable however to have different ratios
of neutrino and antineutrino data. Hence, we also analyze the cases
$(1:2)$ and $(2:1)$. These
can be studied by rescaling the event numbers of the $(1:1)$ mode.

\begin{figure}
\centering

\includegraphics[width=15cm]{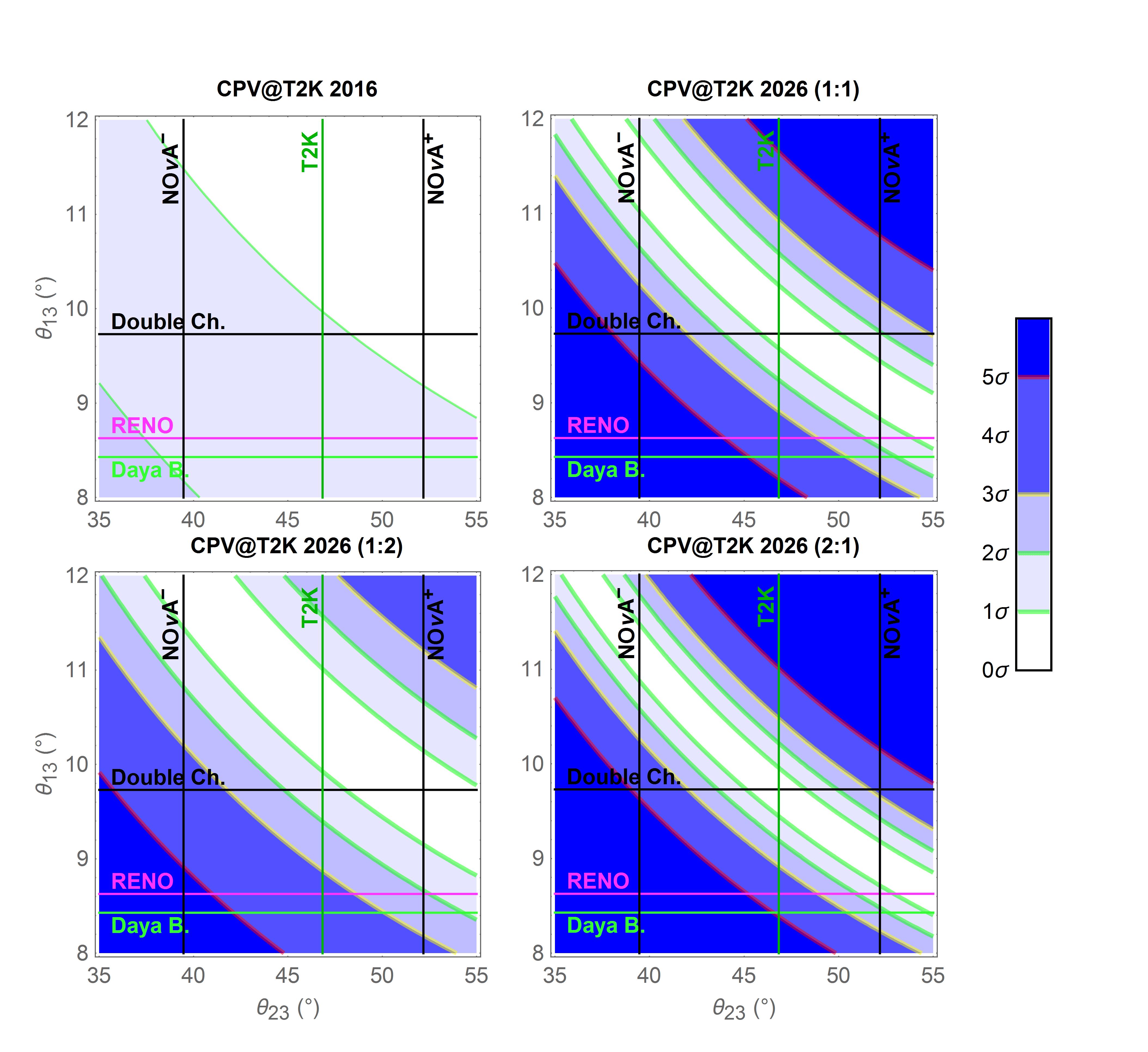}

\caption{CP violation (CPV) in T2K data. {\it Upper left panel:}
  current significance of CPV in T2K (data collected until May 2016
  with $1.5\times 10^{21} $ POT \cite{Abe:2017vif}). {\it Other
    panels:}
  future sensitivities of CPV with $20\times 10^{21} $ POT
  \cite{Abe:2016tii} assuming $\delta_{CP} =-\pi/2$.
The ratios [e.g.\ $(1:2)$, $(2:1)$] denote
  ratios of running times in neutrino and antineutrino modes.\label{fig:CPV}}
\end{figure}

The results are presented in Fig.~\ref{fig:CPV}, where we show the
significance of CPV in both the current data (T2K 2016) and the future
data (T2K 2026) with three different exposure ratios, $(1:1)$, $(1:2)$
and $(2:1)$. As we can see, compared to the T2K 2016 data, the future
data will have more enhanced sensitivity on CPV. Changing the
exposure ratio can affect the result. However whether this increases
or decreases the significance of CPV depends on values of
$\theta_{13}$ and $\theta_{23}$. For example, if
$\theta_{13}=8.43^{\circ}$ and $\theta_{23}=46.8^{\circ}$, which are
the best-fit values of Daya Bay and T2K disappearance measurements,
then the $(2:1)$ mode could reach $4.8\sigma$ significance of CPV
(cf.\ Tab.~\ref{tab:CPV}), better than the $(1:1)$ and $(1:2)$ modes
which have $4.4\sigma$ and $3.9\sigma$ respectively. If $\theta_{23}$
is changed to the best-fit value of NO$\nu$A$^+$,
($\theta_{23}=52.5^{\circ}$), then the $(1:2)$ mode could reach
$2.5\sigma$, larger than $2.3\sigma$ and $2.2\sigma$ in the $(1:1)$ and $(2:1)$
mode.

At first sight it may be surprising that more neutrino data is helping the
significance of CP violation, rather than an equal share of more antineutrino data.
 The reason is that the antineutrino channel has a lower event
rate (about 1/5 of the neutrino channel) and a higher background (about 1:1 signal-to-background
ratio, 4 times higher than the neutrino channel) \cite{Abe:2016tii}.

To sum up, there is optimization potential for the CP phase in future
data taking. Given the large theoretical significance and interest of a maximal CP
phase, this could be worth exploring further.
We note that since $\delta_{CP}$ is expected to receive the largest model corrections
to any neutrino oscillation parameter \cite{Rodejohann:2015nva}, a
value close to a special one like $-\pi/2$ is a strong hint that a
symmetry protects this special value, and thus truly a worthwhile
measurement.

\begin{table}[t]
\centering

\caption{\label{tab:CPV}Significance of CP violation in current and future T2K. The four numbers in each bracket stand for the significance (in the unit of standard deviation $\sigma$) of CPV in the data of T2K2016, T2K2026 $(1:1)$, T2K2026 $(1:2)$ and T2K2026 $(2:1)$, respectively, assuming that the true values of $\theta_{23}$ and $\theta_{13}$ are fixed at different best-fit values.}

\begin{tabular*}{16cm}{@{\extracolsep{\fill}}lccc}
\hline
 & $\theta_{23}=39.5^{\circ}$ (NO$\nu$A${}^-$)  & $\theta_{23}=46.8^{\circ}$ (T2K) & $\theta_{23}=52.2^{\circ}$ (NO$\nu$A${}^+$)\tabularnewline
\hline
$\theta_{13}=9.73^{\circ}$ (Double Ch.) & (1.6, 4.3, 3.8, 4.7) & (1.1, 0.5, 1.4, 0.1) & (0.7, 1.9, 0.2, 3.3)\tabularnewline
$\theta_{13}=8.62^{\circ}$  (RENO) & (1.9, 6.7, 5.4, 7.8) & (1.5, 3.8, 3.5, 4.1) & (1.3, 1.7, 2.1, 1.4)\tabularnewline
$\theta_{13}=8.43^{\circ}$  (Daya B.) & (1.9, 7.0, 5.6, 8.2) & (1.6, 4.4, 3.9, 4.8) & (1.3, 2.3, 2.5, 2.2)\tabularnewline
\hline
\end{tabular*}
\end{table}

\subsection{\label{sec:44}The mass ordering}
\begin{figure}
\centering

\includegraphics[width=12cm]{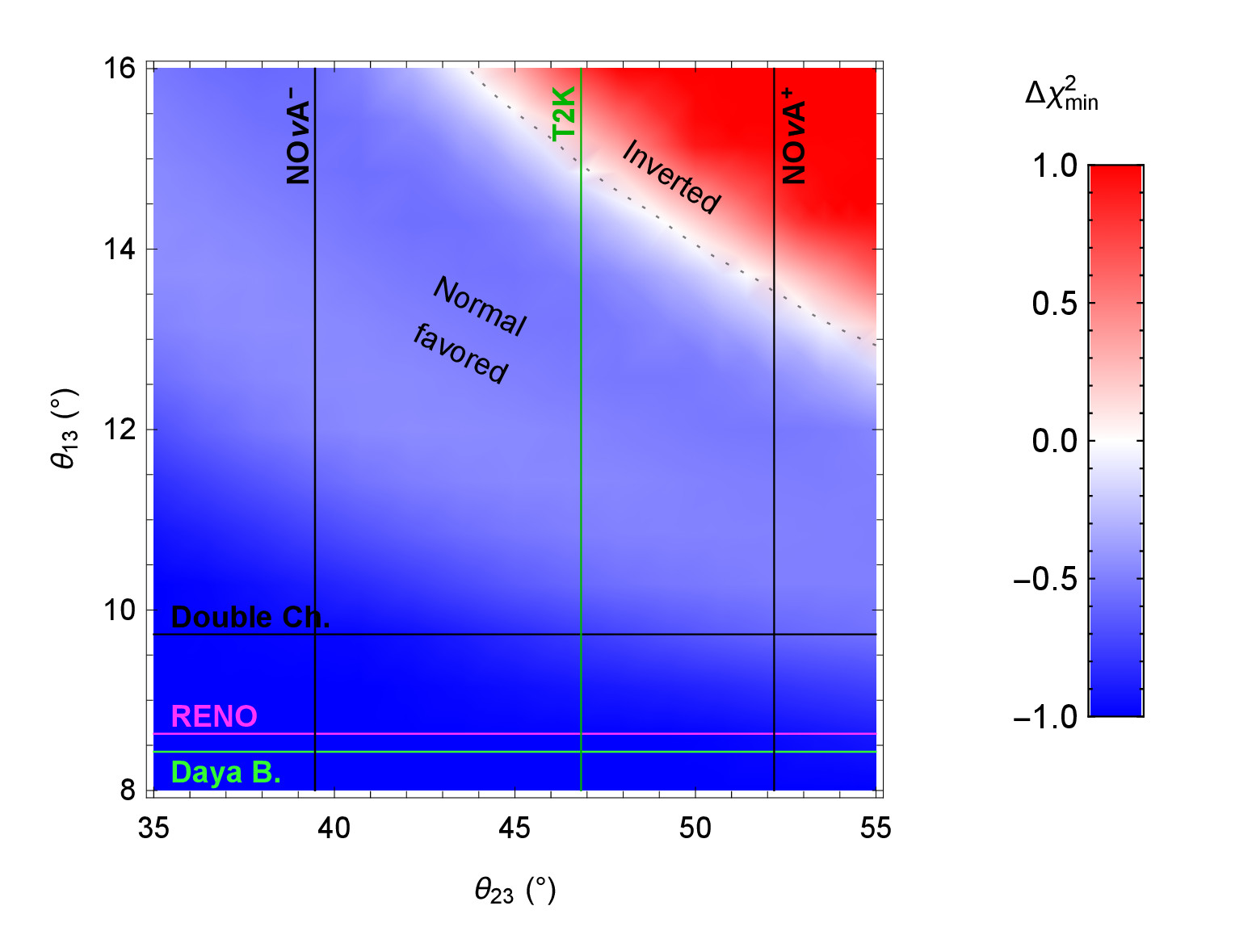}

\caption{Comparison of normal and inverted mass ordering
by $\Delta\chi_{\min}^{2}\equiv\chi_{{\rm min\thinspace N}}^{2}-\chi_{{\rm min\thinspace I}}^{2}$ from
T2K appearance data.
The black dotted curve corresponds to $\Delta\chi_{\min}^{2}=0$. With the one-parameter fitting
performed here, $\Delta\chi_{\min}^{2} = \sigma^{2}$.
 \label{fig:NHIH}}
\end{figure}

Finally, we would like to study the current sensitivity on the mass ordering.
Here the much harder to analyze atmospheric neutrino data has large impact\footnote{We note that
completely independent cosmology data has, depending on used data sets, also some
preference for the normal ordering, see e.g.\  \cite{Hannestad:2016fog}.}, our focus lies
however just on $\nu_{e}$ and $\overline{\nu}_{e}$ appearance data. In the not so far future,
ORCA \cite{Adrian-Martinez:2016fdl} and PINGU \cite{Aartsen:2014oha}
are also expected to contribute to the issue, and JUNO
\cite{An:2015jdp} will settle the question on the mass ordering with a very different method than
the one based on matter effects.

We compare the normal and inverted cases by their minimal
$\chi^{2}$-values,
\begin{equation}
\Delta\chi_{\min}^{2}\equiv\chi_{{\rm min\thinspace N}}^{2}-\chi_{{\rm min\thinspace I}}^{2}\thinspace,\label{eq:eff-39}
\end{equation}
where the subscript ``N'' or ``I'' stands for normal or inverted
ordering respectively; ``min'' stands for the minimization over
the unknown parameter $\delta_{CP}$. More explicitly, $\chi_{{\rm min}}^{2}$
above is computed by
\[
\chi_{\min}^{2}\left(\theta_{23},\thinspace\theta_{13}\right)=\min_{\delta_{CP}}\left[\chi_{{\rm \nu_{e}}}^{2}\left(\theta_{23},\thinspace\theta_{13},\thinspace\delta_{CP}\right)+\chi_{{\rm \overline{\nu}_{e}}}^{2}\left(\theta_{23},\thinspace\theta_{13},\thinspace\delta_{CP}\right)\right],
\]
for any given values of $\theta_{23}$ and $\theta_{13}$. Here $\chi_{{\rm \nu_{e}}}^{2}$
and $\chi_{{\rm \overline{\nu}_{e}}}^{2}$ are constraints from the
T2K $\nu_{e}$ and $\overline{\nu}_{e}$ appearance data respectively,
computed according to Tab.~\ref{tab:fff} and Tab~\ref{tab:nbk}.
The result is shown in Fig.~\ref{fig:NHIH}, where the blue region
has $\chi_{{\rm min\thinspace N}}^{2}<\chi_{{\rm min\thinspace
    I}}^{2}$,
which means NO is favored over IO by the T2K appearance data
while in the red region IO is more favored. We also show the boundary
where $\Delta\chi_{\min}^{2}=0$ by the black dotted curve. Fig.~\ref{fig:NHIH}
implies that except for large $\theta_{13}$ and $\theta_{23}$, the
T2K appearance data prefer NO over IO. This can also be confirmed
in Fig.~\ref{fig:combine13} or Fig.~\ref{fig:combine23}, where
the overlap of the orange and blue regions becomes smaller when the
mass ordering is changed from NO to IO.

%We see from Fig.\ \ref{fig:NHIH}, however, that within the expected values of
%$\theta_{23}$ and $\theta_{13}$ the significance of any mass ordering
%can vary by about one standard deviation.

\section{Conclusion\label{sec:Conclusion}}

Some attention has recently been cast on hints for a nontrivial CP phase and the normal
mass ordering. Since the combination of reactor and accelerator neutrino measurements is the
origin of this we revisit the crucial parameter correlations in oscillation probabilities.
Noting the different available central values of $\theta_{13}$ and $\theta_{23}$ from
the different reactor and long-baseline experiments, we performed an analysis on their
impact of the ranges of the CP phase and the preference of the mass ordering.
While surely not a global fit including all available data, some new insights have been
obtained.

To facilitate comparison of reactor and accelerator data, we have proposed
the effective $\theta_{13}^{\rm eff}$, which is defined in Eq.~(\ref{eq:eff-8}).
The effective $\theta_{13}^{\rm eff}$ can be applied to phenomenological studies
of reactor and accelerator neutrino measurements.
We have demonstrated that the effective $\theta_{13}^{\rm eff}$ can
be used to discuss several correlations of the neutrino parameters including $\delta_{CP}$, $\theta_{13}$ and $\theta_{23}$
by the T2K collaboration very well. For example, we can analytically answer the question why and when the best-fit value of $\delta_{CP}$ tends to be maximal.
It can furthermore straightforwardly explain why
future long-baseline experiments will be more sensitive to the CP phase when
$\delta_{CP}$ is around zero, rather than when it is around $-\pi/2$.
All the analytical arguments are also numerically studied and verified in this paper.

Optimization
potential for the determination of $\delta_{CP}$ around the theoretically
highly interesting value $-\pi/2$ is discussed.
Moreover, using the different available best-fit values
of $\theta_{13}$ and $\theta_{23}$ from reactor experiments and from T2K/NO$\nu$A, respectively,
we studied their impact on the current ranges of $\delta_{CP}$ and the mass ordering.
We have shown that, depending on the true values of $\theta_{13}$ and $\theta_{23}$,
the current sensitivities are subject to change.\\

We hope that our study prompts some discussion on the robustness of
existing hints (or any future hints) in the data, and contributes to future discussion of
neutrino oscillation results.

\appendix

\section{
The effective $\theta_{13}$ approximation
\label{sec:Evaluate-f}}

In this appendix we derive the effective $\theta_{13}^{\rm eff}$ including the
experiment-dependent values $f_s$, $f_c$ and $f_2$, see Eq.\ (\ref{eq:eff-8}).
We start by writing the observable event rate in a neutrino experiment
as follows:
\begin{equation}
\frac{dN}{d\Omega_{f}}=\Delta t\int D(E_{\nu},\thinspace\Omega_{f})P(E_{\nu})\Phi(E_{\nu})dE_{\nu},\label{eq:eff-13}
\end{equation}
Here $\Omega_{f}$ includes final states of all observable particles
after scattering (e.g.\ energy/momentum of an electron), $\Delta t$
is the time of exposure, and ($\Phi$, $P$, $D$) represent  functions describing
neutrino production, propagation and detection, respectively.
More explicitly, $\Phi$ is the flux of neutrinos produced at the
source,  $P$ is the oscillation probability during propagation, and
$D$ is defined as the probability of an incoming neutrino causing
an event in which the final states are given by $\Omega_{f}$. Basically,
$D$ is the differential cross section but practically it should also
include the efficiency of detecting final state particles (e.g.\ detecting
photons in PMTs).

Integrating with respect to $\Omega_{f}$ in Eq.~(\ref{eq:eff-13}),
one can obtain the total event number. To understand the contribution
of each term in Eq.~(\ref{eq:eff}) we decompose it into four terms:
\begin{equation}
P=P_{0}+P_{s}+P_{c}+P_{2}.\label{eq:eff-15}
\end{equation}
Here the four terms are
\begin{eqnarray}
P_{0}(E_{\nu}) & \equiv & 4s_{13}^{2}c_{13}^{2}s_{23}^{2}\,p_{0}(E_{\nu}),\label{eq:eff-17}\\
P_{s}(E_{\nu}) & \equiv & -8\alpha J_{CP}\, p_{s}(E_{\nu}),\label{eq:eff-18}\\
P_{c}(E_{\nu}) & \equiv & 8\alpha J_{CP}\cot\delta_{CP} \, p_{c}(E_{\nu}),\label{eq:eff-19}\\
P_{2}(E_{\nu}) & \equiv & 4\alpha^{2}s_{12}^{2}c_{12}^{2}c_{23}^{2}\,p_{2}(E_{\nu}),\label{eq:eff-20}
\end{eqnarray}
where the energy-dependent parts are defined as
\begin{equation}
p_{0}\equiv\frac{\sin^{2}(1-A)\Delta}{(1-A)^{2}},\label{eq:eff-3-1}
\end{equation}
\begin{equation}
p_{s}\equiv\sin\Delta\frac{\sin A\Delta}{A}\frac{\sin(1-A)\Delta}{1-A},\label{eq:eff-4-1}
\end{equation}
\begin{equation}
p_{c}\equiv\cos\Delta\frac{\sin A\Delta}{A}\frac{\sin(1-A)\Delta}{1-A},\label{eq:eff-5-1}
\end{equation}
\begin{equation}
p_{2}\equiv\frac{\sin^{2}A\Delta}{A^{2}}.\label{eq:eff-6-1}
\end{equation}
In the integral, we are only concerned about the energy-dependent
parts of $P$ so we define
\begin{equation}
N_{X}\equiv\Delta t\int\left(\int Dd\Omega_{f}\right)p_{X}\Phi dE_{\nu},\ {\rm for}\ X=0,\thinspace s,\thinspace c,\thinspace2.\label{eq:eff-14}
\end{equation}
Then the total event number will be
\begin{equation}
N_{{\rm tot}}=s_{23}^{2}\sin^{2}2\theta_{13}N_{0}-8\alpha J_{CP}N_{s}+8\alpha\frac{J_{CP}}{\tan\delta_{CP}}N_{c}+\alpha^{2}\sin^{2}2\theta_{12}c_{23}^{2}N_{2}.\label{eq:eff-21}
\end{equation}
As we can see from Eq.~(\ref{eq:eff-21}), by introducing $N_{0,\thinspace s,\thinspace c,\thinspace2}$,
the dependence of the total event number on the PMNS parameters is
explicitly kept. The $f$-factors introduced in the effective
$\theta_{13}^{\rm eff}$ in Eq.~(\ref{eq:eff-8}) are defined as
\begin{equation}
(f_{s},\thinspace f_{c},\thinspace f_{2})\equiv\frac{(N_{s},\thinspace N_{c},\thinspace N_{2})}{N_{0}},\label{eq:eff-22}
\end{equation}
which shows that they are the contributions of the different oscillation modes $p_{s}$,
$p_{c}$ and $p_{2}$ to the total event number, in comparison to
the dominant mode $p_{0}$. Since Eq.~(\ref{eq:eff-22}) is a ratio,
all overall factors (such as $\Delta t$) will be canceled in
Eq.~(\ref{eq:eff-22}). In practice the $f$-factors can thus be evaluated by
\begin{equation}
f_{X}=\frac{\int F(E_{\nu})p_{X}(E_{\nu})dE_{\nu}}{\int F(E_{\nu})p_{0}(E_{\nu})dE_{\nu}},\ {\rm for}\ X=s,\thinspace c,\thinspace2,\label{eq:eff-23}
\end{equation}
where
\begin{equation}
F(E_{\nu})\equiv\left(\int D(E_{\nu},\thinspace\Omega_{f})d\Omega_{f}\right)\Phi(E_{\nu}).\label{eq:eff-24}
\end{equation}
If one disregards the efficiency of detecting final state particles
(assuming it is $100\%$ or an energy-independent constant), then
the integral in Eq.~(\ref{eq:eff-24}) can be replaced with the total
cross section $\sigma(E_{\nu})$. Therefore we have
\begin{equation}
F(E_{\nu})\propto\sigma(E_{\nu})\Phi(E_{\nu}).\label{eq:eff-25}
\end{equation}

\begin{figure}
\centering

\includegraphics[width=7.5cm]{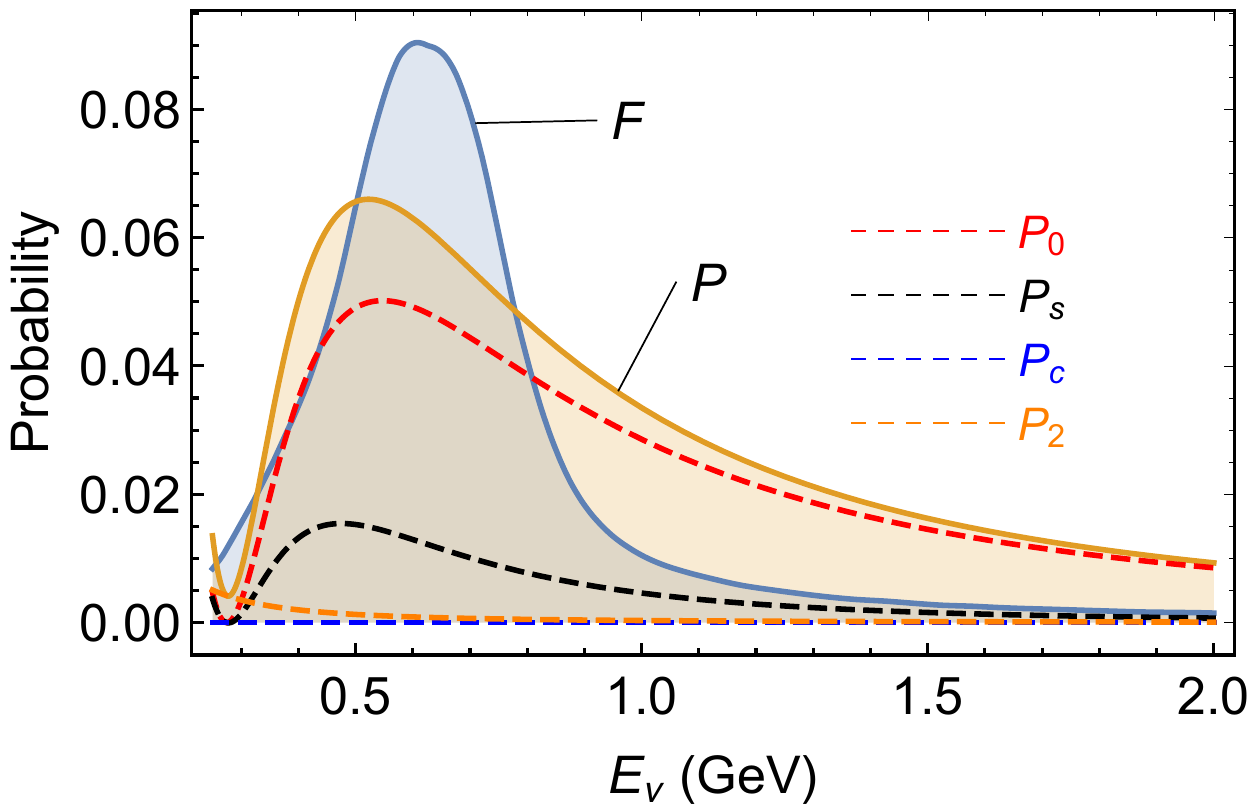}\hspace{1cm}\includegraphics[width=7.5cm]{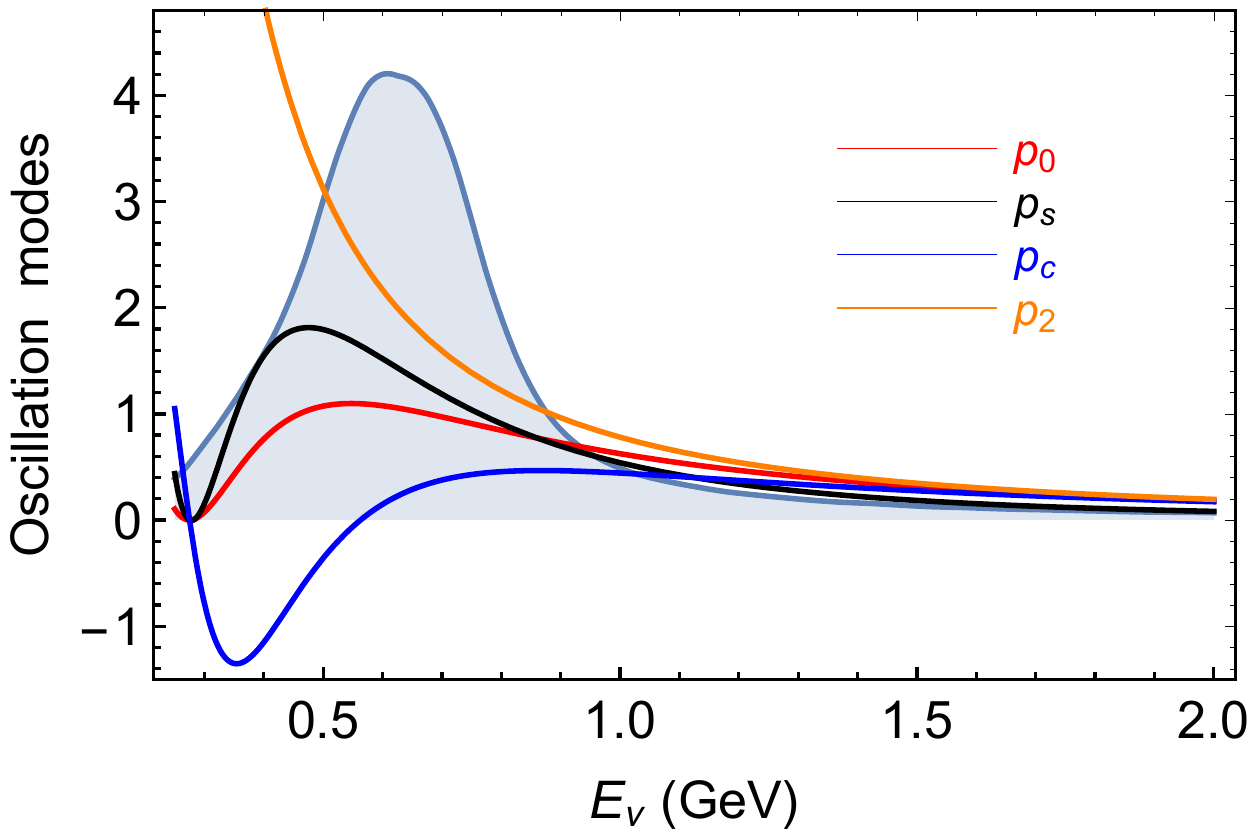}

\caption{The probability of $\nu_{\mu}\rightarrow\nu_{e}$ transitions
and $F(E_{\nu})$ in T2K {[}cf.\ Eqs.~(\ref{eq:eff-15}) and (\ref{eq:eff-25}){]}.
The definitions of $P_{X}$ and $p_{X}$ ($X=0,\thinspace s,\thinspace c,\thinspace2$)
are given in Eqs.~(\ref{eq:eff-17}) to (\ref{eq:eff-6-1}), $P$ is the total
oscillation probability for $(\theta_{23},\,\theta_{13},\,\theta_{12})$=$(45^{\circ}, 8.8^{\circ}, 34^{\circ})$, $\Delta m_{13}^2=2.44\times10^{-3}\, \rm{eV}^2$, $\alpha=0.31$,
%which the best-fit values from table
%\ref{tab:}
and $\delta_{CP} = -\pi/2$ (hence $P_c = 0$).
The function $F$ is presented in arbitrary units.\label{fig:DPF}}
\end{figure}

So far the discussion was general. Let us take now the $\nu_{\mu}\rightarrow\nu_{e}$
measurement of T2K as an example and evaluate the $f$-factors. The T2K experiment uses
an off-axis beam of muon neutrinos generated at the J-PARC accelerator in Tokai,
Japan. For $\Phi(E_{\nu})$ in Eq.~(\ref{eq:eff-25}), we take the
$\nu_{\mu}$ flux from Ref.\ \cite{T2K2013PRD}. The  Super-Kamiokande (SK) far detector
(295 km away from the source)
is a water Cherenkov detector in which neutrinos are detected via
Cherenkov light of charged particles emitted from neutrino scattering.
Since the selected $\nu_{e}$ events are expected from charged current
quasi-elastic (CCQE) scattering, we only need the CCQE cross section
for $\sigma(E_{\nu})$ in Eq.~(\ref{eq:eff-25}). In water ($H_{2}O$,
where we only consider $^{16}O$) we use the
CCQE cross section data from GENIE \cite{Andreopoulos:2009rq}.
Given the cross section and the neutrino flux, we can compute $F(E_{\nu})$
in Eq.~(\ref{eq:eff-25}), up to an irrelevant normalization factor.
The shape of $F(E_{\nu})$ is presented in Fig.\ \ref{fig:DPF}.

Next we compute the integrals in Eq.~(\ref{eq:eff-23}) with
$G_{F}N_{e}=7.01\times10^{-14}\, {\rm eV}$ (corresponding to the terrestrial matter density
$2.6\, {\rm g/cm^{3}}$, or proton density $1.3\, {\rm g/cm^{3}}$ ) and $|\Delta m_{31}^{2}|=2.44\times10^{-3}\,{\rm eV}^{2}$. The numerical
results are listed in Tab.~\ref{tab:fff}, including both $\nu_{e}$
and $\overline{\nu}_{e}$ modes, NO and IO.  For IO, one should use
negative $\Delta m_{31}^{2}$ so $\Delta$ and $A$ are negative,
leading to a negative $f_{s}$. For the $\overline{\nu}_{e}$ mode,
one should change the neutrino flux and cross section correspondingly,
and flip the signs of $A$ and $\delta_{CP}$, which means the negative
sign before the second term of Eq.~(\ref{eq:eff}) should change to
positive. Since we want a unified definition of $\theta_{13}^{{\rm eff}}$
for both $\nu_{e}$ and $\overline{\nu}_{e}$ {[}i.e.\ the form of
Eq.~(\ref{eq:eff-13}) should also apply to $\overline{\nu}_{e}${]}, we
add a negative sign in $p_{s}$
for the $\overline{\nu}_{e}$ mode. As a consequence, $f_{s}$ for
$\overline{\nu}_{e}$ with NO or IO is negative or positive, respectively.

Note that $f_{c}^{{\rm T2K}}$ is much smaller than $f_{s}^{{\rm T2K}}$
which implies that at NLO the $CP$-even term in Eq.~(\ref{eq:eff})
has a much smaller contribution to the oscillation probability than
the $CP$-odd term. Let us discuss the physical meaning
of the $f$-factors to understand why $f_{c}^{{\rm T2K}}$ is so small.
As shown in Fig.\ \ref{fig:DPF}, the function $F(E_{\nu})$ which
is proportional to both the detection rate $D$ and the flux $\Phi$,
peaks at about 0.6 GeV but is suppressed at both high and low energies.
This is because if $E_{\nu}$ is too high, the flux drops
exponentially; if $E_{\nu}$ is too low, the cross section suppresses
$F$ (the flux drops as well).  On the other hand, the oscillation
probability $P$ also shows similar dependence on $E_{\nu}$ (if one
only focuses on the dominating first peak of oscillation).
Therefore, in order to obtain sizable event numbers,
it is important to make $F$ (which is essentially flux times cross
section) and the probability $P$ overlap as much as possible when they
obtain their maximal values.
%as much as possible to obtain large event numbers.

In the left panel of Fig.\ \ref{fig:DPF}, we show the contribution
of each term of the probability, i.e.\ $P_{0}$, $P_{s}$, $P_{c}$
and $P_{2}$ {[}cf.\ Eqs.~(\ref{eq:eff-15})-(\ref{eq:eff-20}){]}.
 In the right panel, we focus on the pure oscillation modes, $p_{0}$,
$p_{s}$, $p_{c}$ and $p_{2}$ {[}cf.\ Eqs.~(\ref{eq:eff-3-1})-(\ref{eq:eff-6-1}){]}
which are directly related to the $f$-factors according to Eq.~(\ref{eq:eff-23}).
The physical meaning of $f_{X}$ ($X=s,\thinspace c,\thinspace2$)
is how much the corresponding oscillation mode $p_{X}$ and the function
$F(E_{\nu})$ overlap, which determines the sensitivity of the experiment
on this mode itself. As we can see, $p_{c}$ has the least overlap
with the shaded region corresponding to $F$, and it also contains a negative part which
causes cancellation with the positive part. This explains why $f_{c}$
is much smaller than the other $f$-factors.

Although the smallness of $f_{c}$ is experiment dependent, it should be a general feature in accelerator neutrino experiments because usually the experiments should be designed so that $F(E_{\nu})$ peaks at the same energy where the oscillation mode $\sin^2\Delta$ peaks. The oscillating part in $p_c$ [cf.~(\ref{eq:eff-5-1})]  approximates in the valid limit of small matter effects as
$\sin\Delta\cos\Delta$, which if integrated with the function $F$
(see Fig.\ \ref{fig:DPF})
%in the shape of $\sin^2\Delta$
will always lead to the similar cancellation.
Therefore one can conclude that in general $f_{c}$ is small in such experiments.

\begin{figure}
\centering

\includegraphics[width=8cm]{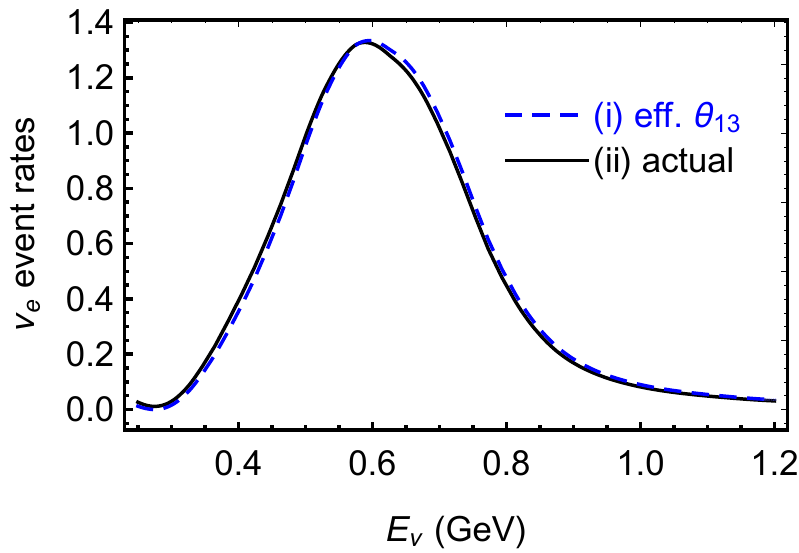}

\caption{Comparison of $\nu_{e}$ event rates (in arbitrary units) computed
(i) from the effective $\theta_{13}^{\rm eff}$ formula {[}cf.\ Eqs.~(\ref{eq:eff-2})
and (\ref{eq:eff-8}){]} and (ii) from the actual oscillation formula {[}cf.\ Eq.~(\ref{eq:eff}){]}
applied to T2K.
%{\bf WR: start plot from 0.3 GeV to avoid confusion.}
\label{fig:eventrates}}
\end{figure}

Finally, we would like to discuss the validity of the effective
$\theta_{13}^{\rm eff}$ when applied to accelerator neutrino experiments. When
introducing the effective $\theta_{13}^{\rm eff}$, we actually assume that
the observable effects of the NLO and NNLO terms are only their contributions
to the total event number, i.e.\ the distortion of the shape of $\nu_{e}$($\overline{\nu}_{e}$)
distribution due to the different oscillation modes is assumed to be negligible.
However, for experiments with very high sensitivities, this assumption might
not hold. To understand the differences between the effective $\theta_{13}$
and the conventional treatment, we compare the event rates in Fig.~\ref{fig:eventrates}.
The event rates are plotted in arbitrary units, computed via $F(E_{\nu})\times P(E_{\nu})$
where $P(E_{\nu})$ takes either the form of Eq.~(\ref{eq:eff-2}) {[}with $\theta_{13}^{{\rm eff}}$
defined in (\ref{eq:eff-8}){]} or of Eq.~(\ref{eq:eff}), plotted by
the blue dashed or the black solid curve, respectively.
From Fig.~\ref{fig:eventrates} we can see that the difference between
the two curves is very small, which implies the approximation adopted
in the effective $\theta_{13}$ could make a difference only in experiments
with very high event numbers (i.e.\ very small statistical uncertainties)
as well as correspondingly very small systematic uncertainties. More
explicitly, if we assume the difference is about $2\%$ as a typical
value indicated in Fig.~\ref{fig:eventrates}, then the total event
number $N_{{\rm tot}}$ should be (according to $1/\sqrt{N_{{\rm tot}}}\sim2\%$)
larger than ${\cal O}(10^{3})$ to distinguish between the two curves.
Current on-going accelerator neutrino experiments such as T2K, MINOS,
NO$\nu$A cannot reach such high statistics while the future experiment
DUNE experiment will do so (861 $\nu_{e}$ events and 167 $\overline{\nu}_{e}$
events for 4 years of running \cite{Acciarri:2015uup}). However, the above requirement
does not take the energy resolution into account (i.e.\ assumes no
uncertainty in the energy measurement). Since the areas under the
two curves are the same (by definition), the energy resolution has
to be very high to see any differences between the two curves. The
horizontal differences are typically around ${\cal O}(1\%)$, which
implies that the energy resolution should reach $\Delta E_{\nu}/E_{\nu}<{\cal O}(1\%)$
to distinguish between them. Therefore, we can draw the conclusion
that the approximation in the effective $\theta_{13}$ should be valid
for experiments with
$N_{{\rm tot}}<{\cal O}(10^{3})$ or $\Delta E_{\nu}/E_{\nu}>{\cal O}(1\%)$.

%{\bf WR: so, can I sum this up by saying simply that
%$\cos\Delta\frac{\sin A\Delta}{A}\frac{\sin(1-A)\Delta}{1-A}$
%is small for $E_\nu =  0.6$ GeV?
%XJ: no, the above paragraph is irrelevant to $\cos\Delta\frac{\sin A\Delta}{A}\frac{\sin(1-A)\Delta}{1-A}$ or .6GeV.
%Are you talking about the paragraph before the above one? That
%paragraph ends with "...why $f_c$ is much smaller than the other
%f-factors." }

\begin{figure}
\centering

\includegraphics[width=8cm]{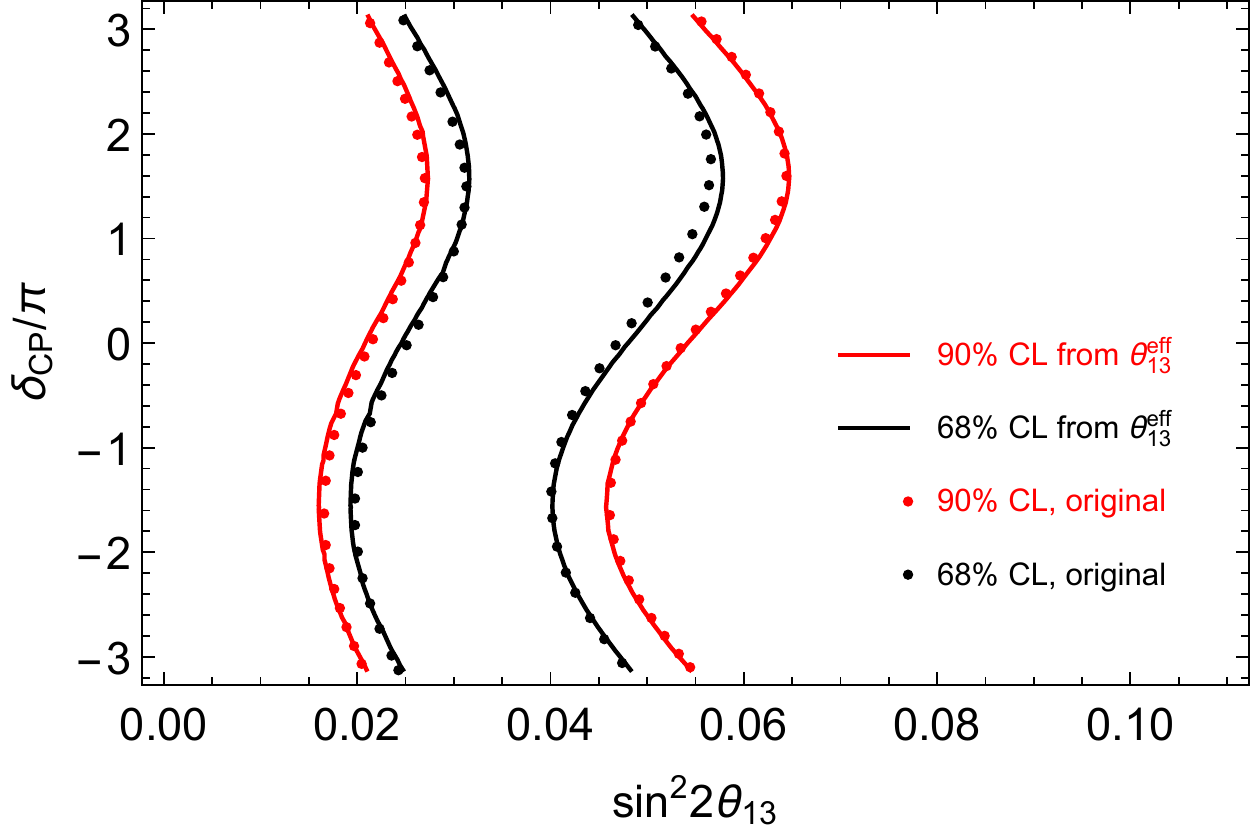}\hspace{2cm}

\caption{Reproduced constraints (red and black curves) on $(\theta_{13},\thinspace\delta_{CP})$
from our effective $\theta_{13}$ framework compared to the original
T2K result \cite{Abe:2017vif} (red and black points) for the normal mass ordering.
The agreement shows that the effective $\theta_{13}$ framework is very accurate
to describe the sensitivity of T2K $\nu_{\mu}\rightarrow\nu_{e}$ data
on $(\theta_{13},\thinspace\delta_{CP})$.   \label{fig:fit-t2k}}
\end{figure}

Using the effective $\theta_{13}$ framework, we can approximately reproduce the constraint of the T2K data on $(\theta_{13},\thinspace\delta_{CP})$, as shown in
Fig.\ \ref{fig:fit-t2k}. Compared with the original
constraint (black and red points) from Ref.\ \cite{Abe:2017vif}, our reproduced result  agrees with it very well. Thisr
 implies that our effective $\theta_{13}$ framework
is able to quantitatively describe the sensitivity of T2K
$\nu_{\mu}\rightarrow\nu_{e}$ data
on $(\theta_{13},\thinspace\delta_{CP})$ with sufficient accuracy.
The details of reproducing the constraint are given next.

Since the expected event number $\mu$ should linearly depend on the
probability of $\nu_{e}$ appearance, it should also linearly depend
on $S^{2}\equiv\sin^{2}2\theta_{13}^{{\rm eff}}$. So we can write
\begin{equation}
\mu=kS^{2}+b,\label{eq:eff-28-1}
\end{equation}
where $k$ is a scale factor and $b$ can be regarded as the background
because it equals the expected event number without neutrino oscillation.
The observed event number $n$ should obey a Poisson distribution, hence
the $\chi^{2}$-function of $\mu$ is
\begin{equation}
\chi^{2}(\mu)=2\left(\mu-n+n\ln\frac{n}{\mu}\right).\label{eq:eff-35}
\end{equation}
By plugging Eq.~(\ref{eq:eff-28-1}) into Eq.~(\ref{eq:eff-35}) we
obtain the $\chi^{2}$-function of $S^{2}$,
\begin{equation}
\chi^{2}(S^{2})=2\left(kS^{2}+b-n+n\ln\frac{n}{kS^{2}+b}\right).\label{eq:eff-27-1}
\end{equation}
In the recent T2K appearance data, the observed event
number $n$ is $32$ ($\nu_{e}$) or $4$ ($\overline{\nu}_{e}$); the two parameters $k$ and $b$ can be determined
by comparing Eq.~(\ref{eq:eff-27-1}) to the original T2K result \cite{Abe:2017vif}, which are listed in Tab.~\ref{tab:nbk}.
%gives $(k,\thinspace b)=(1.86\times10^{2},\thinspace4.84)$.
\begin{table}[h]
\centering

\caption{\label{tab:nbk}The $k$ and $b$ parameters for the T2K data.}

\begin{tabular}{ccccc}
\hline
 & \hspace{0.3cm}$\nu_{e}$ normal\hspace{0.3cm} & \hspace{0.3cm}$\nu_{e}$ inverted\hspace{0.3cm} & \hspace{0.3cm}$\overline{\nu}_{e}$ normal\hspace{0.3cm} & \hspace{0.3cm}$\overline{\nu}_{e}$ inverted\hspace{0.3cm}\tabularnewline
\hline
$n$ & 32 & 32 & 4 & 4\tabularnewline
\rule[0ex]{0pt}{3ex}  $k$ & $1.86\times10^{2}$  & $1.55\times10^{2}$ & $51.5$ & $55.9$\tabularnewline
\rule[0ex]{0pt}{3ex}  $b$ & $4.8$ & $5.4$ & $2.5$ & $2.5$\tabularnewline
\hline
\end{tabular}

\end{table}

%We have here extracted $k$ and $b$ from Fig.\ 39 of Ref.\
%\cite{Abe:2017vif}.
%{\bf WR: from where in that paper? Some details would be good.} {\bf  XJ: The curves in Fig.~\ref{fig:fit-t2k} can be regarded as functions of $k$ and $b$, according to Eq.~(\ref{eq:eff-27-1}). So one can use Eq.~(\ref{eq:eff-27-1}) to fit the dots in Fig.~\ref{fig:fit-t2k} to get the values of $k$ and $b$.}
Eq.~(\ref{eq:eff-27-1}) can be further converted to the
$\chi^{2}$-function of $(\theta_{13},\thinspace\delta_{CP})$  if $\alpha$,
$\theta_{23}$ and $\theta_{12}$ are fixed or marginalized over. In Fig.\ \ref{fig:fit-t2k}
we simply use fixed values $\alpha=0.0307$, $\theta_{23}=46.5^{\circ}$
and $\theta_{12}=33.4^{\circ}$ to generate the black and red curves.
Although the original T2K result was obtained by marginalizing all
nuisance parameters, our result using fixed values already agrees
well with it.

\begin{acknowledgments}
WR is supported by the DFG with grant RO 2516/6-1 in the Heisenberg program.
\end{acknowledgments}

\bibliographystyle{apsrev4-1}
\bibliography{ref}

\end{document}